\documentclass[prx, twocolumn, longbibliography,
amssymb,superscriptaddress]{revtex4-1}
\usepackage{graphicx, bm, amsmath, amsfonts,amssymb}

\usepackage{subfigure}

\def\v#1{\mathbf{#1}}

\def\t#1{\widetilde{#1}}

\def\up{\uparrow}
\def\down{\downarrow}

\begin{document}

\title{Fractionalized time reversal, parity and charge conjugation symmetry in topological superconductor: a possible origin of three generations of neutrinos and mass mixing}

\author{Zheng-Cheng Gu}
\affiliation{ Department of Physics, The Chinese University of Hong Kong, Shatin, New Territories, Hong Kong}
\affiliation{Perimeter Institute for Theoretical Physics, Waterloo, Ontario, N2L2Y5, Canada}

\begin{abstract}
The existence of three generations of neutrinos(charged leptons/quarks) and their mass mixing are deep mysteries of our universe. The history of neutrino physics can be traced back to Majorana's elegant work on a real solution of the Dirac equation -- known as the Majorana fermion.
Recently, Majorana's spirit returns in modern condensed matter physics -- in the context of topological Majorana zero modes in certain classes of topological superconductors(TSCs).
In this paper, we attempt to investigate the topological nature of the neutrino by assuming that a relativistic Majorana fermion can be divided into four topological Majorana zero modes at cutoff energy scale, e.g. planck scale.
We begin with an exactly solvable $1$D lattice model which realizes a $T^2=-1$ time reversal symmetry protected TSC, and show that a pair of topological Majorana zero modes can realize a $T^4=-1$ time reversal symmetry.
Moreover, we find that a pair of topological Majorana zero modes can also realize a $P^4=-1$ parity symmetry and even a nontrivial $\overline C^4=-1$ charge conjugation symmetry.
Next, we argue that the origin of three generations of neutrinos(charged leptons and quarks) can be naturally explained as three distinguishable ways of forming a pair of complex fermions(with opposite spin polarizations) out of four topological Majorana zero modes, characterized by the $T^4=-1$, $(TP)^4=-1$ and $(T \overline C)^4=-1$ fractionalized symmetries that each complex fermion carries at cutoff energy scale.
 Finally, we use a semiclassical approach to compute the neutrino mass mixing matrix at leading order(LO), e.g., in the absence of $CP$ violation correction. We obtain $\theta_{12}=31.7^\circ, \theta_{23}=45^\circ$ and $\theta_{13}=0^\circ$(the golden ratio pattern), which is consistent with an $A_5$ flavor symmetry pattern.
We further predict an exact mass ratio of the three mass eigenstates of neutrinos with $m_1/m_3=m_2/m_3=3/\sqrt{5}$ and an effective mass of neutrinoless double beta decay $m_{0\nu\beta\beta}=m_1/\sqrt{5}$.
\end{abstract}

\maketitle

\section{Introduction}
\subsection{The neutrino puzzles}
The neutrino, first discovered in 1956\cite{neutrino1} and named as the "ghost particle", has extremely weak interactions with other matters, and it is one of the big mysteries to us and has a deep relationship with the physics of early universe. The theoretical perspective of neutrino physics can be traced back to Ettore Majorana's elegant work\cite{Majorana} on a real solution of the Dirac equation -- known as the Majorana fermion. Unfortunately, for over a century, we have found that all the fundamental particles have their own anti-particles and therefore are described as Dirac fermions. However, the neutrino is still possible to be a Majorana fermion because it does not carry electric charge. In the Standard Model(SM), the neutrino is described as a chiral Weyl fermion with zero rest mass\cite{standardmodel}.

A cutting-edge step toward understanding this big puzzle has been taken by the neutrino oscillation experiments during the past decade\cite{neutrinoexp1,neutrinoexp2,neutrinoexp3,neutrinoexp4,neutrinoexp5,neutrinoexp5.5,neutrinoexp6,neutrinoexp7,neutrinoexp8,neutrinoexp9,neutrinoexp10,neutrinoexp11}. These experiments have confirmed that the neutrino has a nonzero mass, at energy scale of $0.1eV$. This big discovery starts to shake the foundation of modern particle physics, which is built on the well tested SM. So far, it is the first and the only new physics beyond the SM that has been observed experimentally. Nevertheless, it is still unclear whether the neutrino is a Dirac fermion or a Majorana fermion. The smoking gun experiment that might be able to distinguish these two cases is the so-called neutrinoless double-$\beta$ decay. Unfortunately, such experimental evidence is still missing so far\cite{doublebeta1,doublebeta2,doublebeta3,doublebeta4}.

The biggest puzzles are: (1)Where does the neutrino mass come from? (2)Why there are three generations of neutrinos(charged-leptons/quarks)\cite{neutrino2}? (3)Where do those mystery mixing angles come from? An elegant way to explain the origin of neutrino mass is to introduce a sterile right-handed neutrino that does not carry any electroweak charge, and through the so called seesaw mechanism\cite{seesaw1,seesaw2,seesaw3,seesaw4} -- by introducing a heavy Majorana mass for the right-handed sterile neutrino, a small mass for the left-handed light neutrino can be induced. Apparently, the seesaw mechanism requires the neutrino to be a Majorana fermion\cite{Majorananeutrino}. Nevertheless, the rest two puzzles have not been solved in a natural and simple way so far.The observed neutrino mass mixing angles clearly indicates certain (approximate) flavor symmetry, but it is unclear where the mystery flavor symmetry comes from.
In this paper, we aim to propose a topological scenario to explain the origin of three generations of neutrinos and the emergence of flavor symmetry.

\subsection{Topological Majorana zero mode and symmetry fractionalization}
After almost $80$ years since Majorana's mystery disappearance, his spirit returns in modern condensed matter physics \cite{WilczekMajorana} -- in the context of topological Majorana zero modes in certain classes of topological superconductors(TSCs)\cite{KitaevMajorana,ReadMajorana}.
A topological Majorana zero mode can be viewed as a "half fermion" and carries Non-Abelian statistics. Searching for topological Majorana zero modes has become a fascinating subject both theoretically\cite{SankaMajorana,ChetanMajorana,PALeeMajorana,LiangMajorana,KaneMajorana}and experimentally. Very recently, experimental evidences for the existence of topological Majorana zero modes in 1D have been observed in superconductor/semiconductor
nanowire devices\cite{Majoranaexp1,Majoranaexp2,Majoranaexp3} based on an elegant theoretical proposal\cite{Majoranatheory1,Majoranatheory2}. The most interesting property of topological Majorana zero modes is that they can carry fractionalized time reversal, parity and charge conjugation symmetry.
In this paper, we will begin with an exactly solvable $1$D condensed matter model which realizes a $T^2=-1$ time reversal symmetry protected TSC and show that the pair of topological Majorana zero modes on its ends realizes a $T^4=-1$ representation of time reversal symmetry. We then show that such kind of fractionalized time reversal symmetry for a pair of topological Majorana zero modes can be generalized into $P^4=-1$ parity symmetry and even a nontrivial $\overline C^4=-1$ charge conjugation symmetry(or particle-hole symmetry in terms of condensed matter physics language) as well. These fractionalized $\overline C, P$ and $T$ symmetries make a topological Majorana zero mode really behave like a "half fermion".

\subsection{A topological aspect of the Majorana fermion}
Despite the similarity in mathematical structure, from a usual perspective, the localized topological objects -- topological Majorana zero modes have nothing to do with the propagating relativistic Majorana fermion.
In this paper, we attempt to investigate the topological nature of neutrinos by establishing a connection between a Majorana fermion and topological Majorana zero modes -- assuming \emph{a relativistic Majorana fermion can be divided into four topological Majorana zero modes at cutoff energy scale, e.g., planck scale.}

Actually, the topological aspect of elementary particles was first proposed by Lord Kelvin $150$ years ago. He conjectured that different atoms could be described as vortex rings linking in topologically distinguishable ways. The attractive point of Kelvin's idea is that the topological objects -- vortex rings, are (topologically) robust elementary objects that can not be further divided. Nowadays, we know that Kelvin was wrong and atoms can be divided into quarks and leptons; however, we can still raise the similar question: what are quarks and leptons made up of? In our scenario, they are all made up of fundamental topological Majorana zero modes, which are the quantum analogs of Kelvin's vortex rings.


\begin{figure}[t]
\begin{center}
\includegraphics[width=6cm]{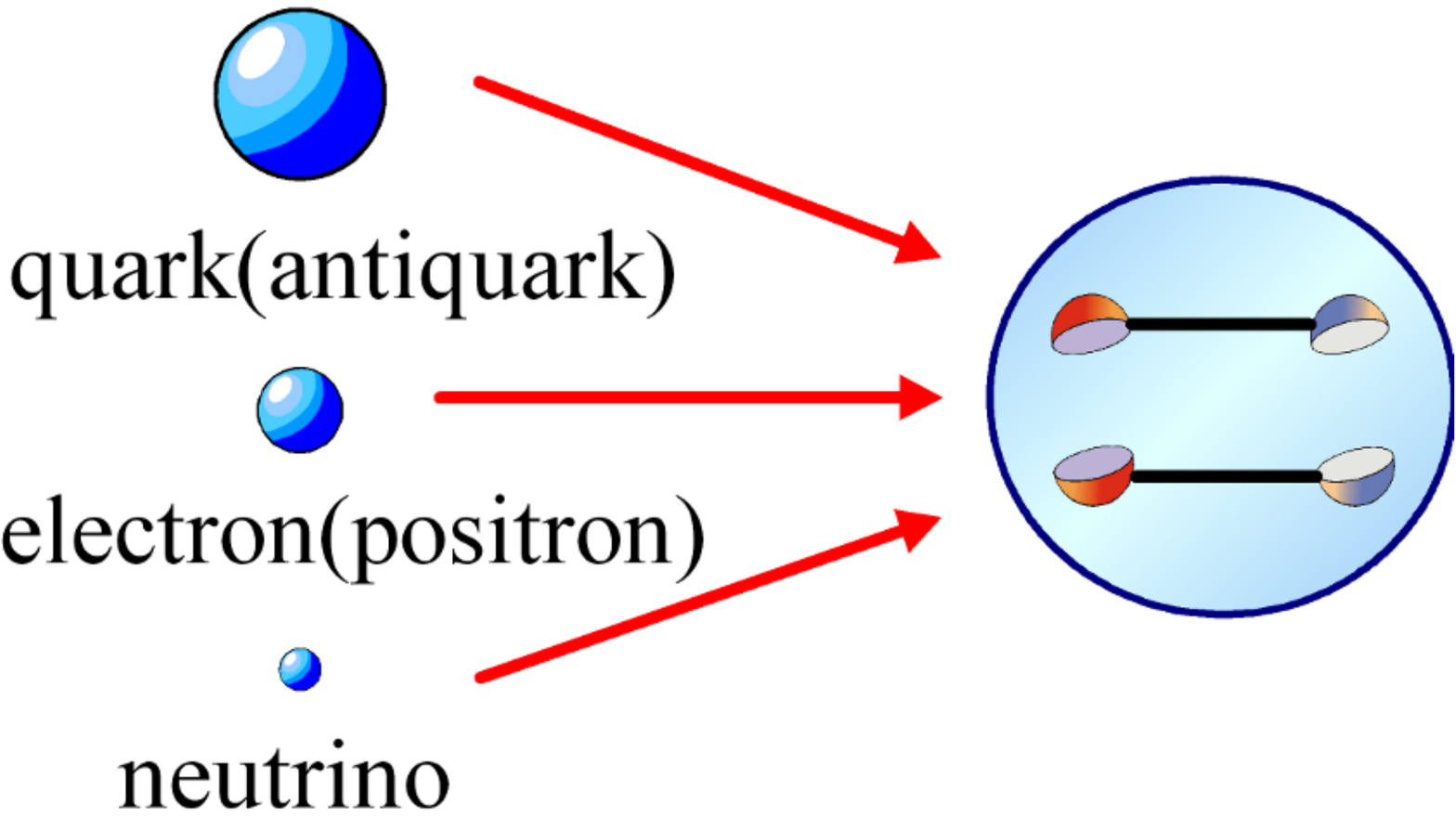}
\caption{(color online)
All fundamental particles, e.g., quarks(antiquarks), electrons(positrons) and neutrinos are described by a two component complex Weyl fermion which is equivalent to a four component real Majorana fermion. A potential underlying TQFT regulation for SM suggests that
a relativistic Majorana fermion can be divided into four topological Majorana zero modes at cutoff energy scale, e.g., planck scale. However, since a single topological Majorana zero mode carries Non-Abelian statistics and can not exist as a point-like particle in 3+1D, it must be attached to the end of a string-like object. Thus, a Majorana fermion consisting of four topological Majorana zero modes should be regarded as two tiny open strings, each of them attached by two topological Majoraan zero modes on its ends.
}\label{string}
\end{center}
\end{figure}

Mathematically, it is well known that 2+1D topological quantum particles, e.g., topological Majorana zero modes, can be described by excitations of an underlying topological quantum field theory(TQFT) or its equivalent algebraic description --- the unitary modular tensor category(UMTC) theory. Therefore, our scenario suggests that an ultimate unification theory for all physical laws at cutoff scale, e.g., planck scale could be described by an underlying 3+1D TQFT which is ultra-violet(UV) complete and differmorphism invariant. Despite the absence of propagating modes in a TQFT, there is no obstruction at planck scale since nothing can propagate at that scale, e.g., light can not escape from a black hole! At low energy, relativistic quantum field theory easily emerges from an ultimate TQFT if certain topological objects are condensed. For example, let us consider a BF theory in 3+1D with a topologically invariant action $S_{\text{top}}=\frac{1}{2\pi}\int B\wedge F$, where $B$ is a two-form gauge field that couples to the string-like objects while $F$ is the field strength of a one-form gauge field that couples to the particle-like objects. In the string condensed phase, $B$ is in the Higgs phase and will acquire a mass term $m^2B^2$. By integrating out the $B$ field, it is easy to see the emergence of Maxwell term $\frac{1}{m^2}F^2$. Unfortunately, so far it is unclear how to derive SM and Einstein gravity from any known TQFT. We believe that the concept of topological Majorana zero mode might play an essential role along this line of thinking.  

Topological Majorana zero mode in 3+1D system was first introduced in a nonlinear sigma model in Ref.\cite{KaneMajorana}. Later, it was pointed out that a single Majorana zero mode carries Non-Abelian statistics and can not consistently exist in 3+1D as a point-like particle\cite{3DTSC}. In fact, it must be attached to the end of a string-like extensive object, as shown in Fig. \ref{string}. Such a composite structure is consistent with the mathematical foundation of 3+1D TQFT, namely, unitary modular tensor 2-category(UMT2C) theory which consists of both point-like and string-like simple objects\cite{KongTQFT}. For example, a topological BF theory in 3+1D contains both particle-like(gauge charge) and string-like(flux line) sources. Therefore, a Majorana fermion consisting of four topological Majorana zero modes should be regarded as two tiny open strings at cutoff scale, e.g. planck scale, and each string is attached by two topological Majorana zero modes on its ends. At low energy, if the string has a very big tension, the pair of topological Majorana zero modes will be confined. In this limit, two tiny strings with four topological Majorana zero modes will just describe two possible particle states with opposite spin polarizations, which is the usual perspective for a Majorana fermion. Finally, we note that our topological scenario can also be applied to quarks and charged leptons, since a Dirac fermion can always be decomposed into two Majorana fermions.

\subsection{Key assumption and basic logic of the paper}
As having been discussed above, the key assumption of this paper is that \emph{a relativistic Majorana fermion can be divided into four topological Majorana zero modes at cutoff energy scale, e.g., planck scale.} By constructing a 3D lattice model formed by topological Majorana zero modes, we show that relativistic Majorana fields(equivalent to two component Weyl fields) can emerge at low energy as a collective motion of Majorana zero modes. In principle, strong interactions among topological Majorana zero modes can even produce the whole SM at low energy and all the fundamental particles can be regarded as collective motions of Majorana zero modes. (Just like strongly correlated electron systems, the low energy excitations are actually collective motions of electrons.)

Most strikingly, such a simple assumption can even naturally explain the origin of three generations of neutrinos(as well as quarks and charged leptons, since a Dirac field can always be decomposed into two Majorana fields), because there are three inequivalent ways to form a pair of complex fermions(with opposite spin polarizations) out of four topological Majorana zero modes, characterized by $T^4=-1$, $(TP)^4=-1$ and $(T\overline C)^4=-1$ fractionalized symmetry that each complex fermion carries at cutoff energy scale. The above physical statement can also be understood as three and only three inequivalent ways of spontaneously breaking $SL(4,R)$(isomorphic to $SO(3,3)$) spacetime symmetry down to $SO(3,1)$ spacetime symmetry(isomorphic to $SL(2,C)$). (We note that due to the topological invariance, the spacetime symmetry at cut-off energy scale can potentially be enlarged to $SL(4,R)$ instead of the usual Lorentz symmetry $SO(3,1)$.)

\subsection{Main results and outline}
Although topological Majorana zero modes can not be directly observed at low energy in SM, these objects can indeed play the role of the right-handed neutrinos in
the seesaw mechanism at cut-off energy scale.  Moreover, if we assume that the nontrivial charge conjugation symmetry is indeed a $\mathbb{Z}_2$  local(gauge) symmetry, the origin of the right-handed neutrino mass can be explained by spontaneous
gauge symmetry breaking through the Higgs mechanism.

By using the minimal coupling principle, we are able to compute the right-handed neutrino mass mixing matrix with no fitting parameters by using a semi-classical approach and
the obtained mixing angles are consistent with the phemomelogically proposed golden ratio(GR) pattern \cite{masssymmetry1,masssymmetry2,masssymmetry3,masssymmetry4}, which arises from an underlying $A_5$ flavor symmetry. 

At leading order(LO), e.g., neglecting the CP violation corrections, if we further impose the $A_5$ flavor symmetry for the whole lepton sector, we will be able to predict that the left-handed light neutrinos have an inverted hierarchy structure with a special mass ratio $m_1=m_2=\frac{3}{\sqrt{5}}m_3$ and the effective mass of neutrinoless double beta decay $m_{0\nu\beta\beta}=m_1/\sqrt{5}$. In fact, our topological scenario also gives rise to the possible origin of the (approximate) $A_5$ flavor symmetry in extended SM. This is because $SO(3,1))\otimes A_5$ is a subgroup of the enhanced spacetime symmetry $SL(4,R)$. In other words, the $A_5$ flavor symmetry could be part of the spacetime symmetry at cut-off energy scale! 

 Based on the current experimental data for $\Delta m^2_{23}$, within LO approximation, we obtain $m_1=m_2\simeq0.075eV$, $m_3\simeq0.054eV$ and $m_{0\nu\beta\beta}=0.0335eV$. 
Our prediction of (approximate) neutrino masses is also consistent with the current cosmological bound on neutrino masses, where $m_1+m_2+m_3<0.3eV$\cite{cosneutrino}. 

The rest of the paper is organized as follows: In section~\ref{sec: time reversal}, we begin with a $1$D TSC protected by $T^2=-1$ symmetry and show why a pair of topological Majorana zero modes on each end must carry a $T^4=-1$ symmetry. In section~\ref{sec: emergent spin}, we discuss how to realize topological Majorana zero modes and $T^4=-1$ time reversal symmetry in higher dimensions.  In section \ref{appemergent}, we construct a lattice model and argue that the proliferation of topological Majorana zero modes in TSC will lead to the emergence of relativistic dispersion and SU(2) spin. In section \ref{sec: superCPT}, we show that the $\overline C,P,T$ symmetries for a Majorana fermion(assuming it is made up of four topological Majorana zero modes) form a super algebra. In section \ref{sec: field theory}, we generalize the $\overline CPT$ super algebra into relativistic quantum field theory and discuss the origin of (right-handed) neutrino masses. In section \ref{sec: three generation}, we give a simple physical picture for the origin of three generations of neutrinos based on the fractionalized $\overline C,P,T$ symmetries. We further discuss how to use a potential TQFT framework to understand the origin of three generations of neutrinos from spontaneously breaking $SL(4,R)$ spacetime symmetry down to $SO(3,1)$ spacetime symmetry at cutoff energy scale, e.g., planck scale.  
In section \ref{sec: mass mixing}, we derive the right-handed neutrino mass mixing matrix within LO approximation without fitting parameters, which is consistent with an $A_5$ flavor symmetry pattern.
Finally, we summarize the new concepts proposed in this paper and discuss other possible new physics along this direction.

\section{Topological Majorana zero modes and $T^4=-1$ time reversal symmetry}
\label{sec: time reversal}
\subsection{$1$D Majorana chain with $T^2=-1$ time reversal symmetry}
\begin{figure}[t]
\vskip -1.0cm
\begin{center}
\includegraphics[width=8cm]{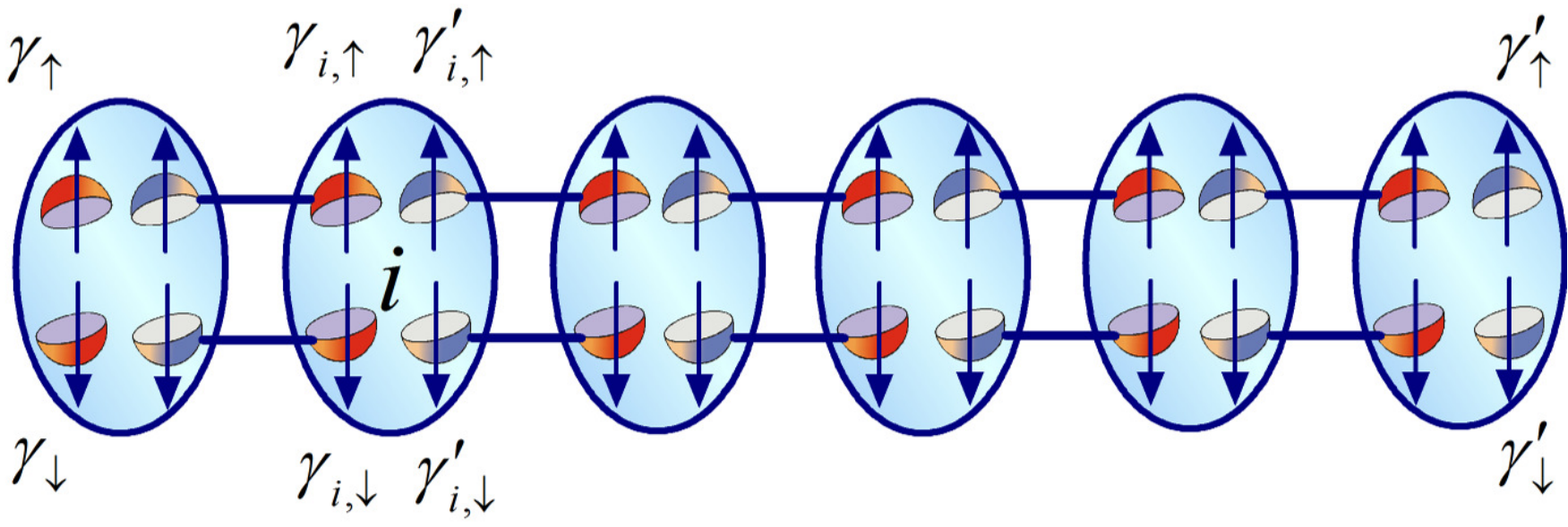}
\vskip -1.0cm
\caption{(color online)A $1$D topological superconductor protected by the $T^2=-1$ time reversal symmetry can be constructed by two copies of Kitaev's Majorana chain with opposite spin species.
We note that each physical site consists of four topological Majorana modes, or two Majorana spinons. The dangling Majorana spinons on both ends become zero modes protected by the time reversal symmetry.
}\label{chain}
\end{center}
\end{figure}

To begin, we consider a $1$D topological superconductor protected by the time reversal symmetry $T^2=-1$, which realizes a special symmetry protected topological(SPT) phase\cite{GuSPT} in $1$D. Literally, such a $1$D TSC has been originally proposed in a $1$D free fermion system with a $T^2=-1$ symmetry(the $\rm{DIII}$ class)\cite{Ryuperiod,Kitaevperiod}.
The simplest model that realizes such a $1$D topological superconductor is just two copies of Kitaev's Majorana chain\cite{KitaevMajorana} with opposite spin species, as seen in Fig.\ref{chain}, described by the following Hamiltonian:
\begin{eqnarray}
H=\sum_{i=1}^N\sum_{\sigma} i\sigma\gamma_{i,\sigma}^{\prime}\gamma_{i+1,\sigma},\label{H}
\end{eqnarray}
The Majorana operators $\gamma_{i,\sigma}$ and $\gamma_{i,\sigma}^\prime$ satisfy:
\begin{eqnarray}
\{\gamma_{i,\sigma},\gamma_{i^\prime,\sigma^\prime}^\prime\}=0; \quad \{\gamma_{i,\sigma},\gamma_{i^\prime,\sigma^\prime}\}=2\delta_{ii^\prime}\delta_{\sigma\sigma^\prime}\label{commutate}
\end{eqnarray}
In terms of the complex fermion operators:
\begin{eqnarray}
c_{i,\up}=\frac{1}{2}(\gamma_{i,\up}+i\gamma_{i,\up}^\prime);\quad c_{i,\down}=\frac{1}{2}(\gamma_{i,\down}-i\gamma_{i,\down}^\prime)\label{def}
\end{eqnarray}
We can rewrite the above Hamiltonian as:
\begin{eqnarray}
H=\sum_{i=1}^N\sum_{\sigma} \left(c_{i,\sigma}-c_{i,\sigma}^\dagger \right)\left(c_{i+1,\sigma}+c_{i+1,\sigma}^\dagger \right)
\end{eqnarray}
Under the time reversal symmetry, the bulk complex fermion operators transform as usual:
\begin{eqnarray}
T c_{i,\uparrow} T^{-1}&=&
-c_{i,\downarrow}; \quad
T c_{i,\downarrow} T^{-1}= c_{i,\uparrow} \nonumber\\ T c_{i,\uparrow}^\dagger T^{-1}&=&
-c_{i,\downarrow}^\dagger; \quad
T c_{i,\downarrow}^\dagger T^{-1}= c_{i,\uparrow}^\dagger,
\end{eqnarray}
According to Eq.(\ref{def}), it is clear that Majorana spinons $(\gamma_{i,\up},\gamma_{i,\down})$ and $(\gamma_{i,\up}^\prime,\gamma_{i,\down}^\prime)$ on a single site should transform in the same way:
\begin{eqnarray}
 T \gamma_{i,\uparrow} T^{-1}&=&
-\gamma_{i,\downarrow}; \quad
T \gamma_{i,\downarrow} T^{-1}= \gamma_{i,\uparrow} \nonumber\\
 \quad T \gamma_{i,\uparrow}^\prime T^{-1}&=&
-\gamma_{i,\downarrow}^\prime; \quad
T \gamma_{i,\downarrow}^\prime T^{-1}= \gamma_{i,\uparrow}^\prime
\label{T}
\end{eqnarray}

Although the model Hamiltonian Eq.(\ref{H}) is very simple, it describes a nontrivial time reversal symmetry protected TSC, characterized by topological Majorana zero modes and the symmetry fractionalization on its ends. On the other hand, Eq.(\ref{H}) also describes a fixed point Hamiltonian with zero correlation length, therefore all its nontrivial topological properties could be applied to generic models describing the same SPT phase.

As seen in Fig. \ref{chain}, a pair of dangling topological Majorana modes with opposite spins( $\gamma_{\uparrow}\equiv\gamma_{1,\uparrow},\gamma_{\downarrow}\equiv\gamma_{1,\downarrow}$ for left end and
$\gamma_{\uparrow}^\prime\equiv\gamma_{N,\uparrow}^\prime,\gamma_{\downarrow}^\prime\equiv\gamma_{N,\downarrow}^\prime$ for right end) form a Majorana spinon on each end, and Eq.(\ref{T}) implies that the fermion mass term
$i\gamma_{\uparrow}\gamma_{\downarrow}$($i\gamma_{\uparrow}^\prime\gamma_{\downarrow}^\prime$) changes sign under the time reversal. Thus, the pair of topological Majorana modes is stable against $T$-preserving interactions and the Hamiltonian Eq.(\ref{H}) describes a time reversal symmetry protected TSC.
Recent progress on the classification of $1D$ SPT phases\cite{XieSPT1,XieSPT2} further pointed out that the edge topological Majorana zero modes indeed carry the $T^4=-1$ projective representation of time reversal symmetry, rather than the usual $T^2=-1$ representation.
A simple reason why we need such a $T^4=-1$ representation can be explained as following: If we assume a Majorana spinon carries the same $T^2=-1$ representation as Kramers doublets, the total time reversal symmetry action on a single physical site will carry a $T^2=1$ representation as it contains two Majorana spinons. Therefore, a $T^2=-1$ representation for the complex spinon on a single physical site prohibits the same $T^2=-1$ representation for a Majorana spinon.

To understand the origin of the $T^4=-1$ representation, we need to investigate the precise meaning of $T^2=-1$ time reversal symmetry for interacting fermion systems.
Indeed, the local Hilbert space on a single site for the above $T^2=-1$ TSC is a Fock-space which involves both fermion parity odd states $c_{i,\up}^\dagger |0\rangle,c_{i,\down}^\dagger|0\rangle$ and parity even states $|0\rangle,c_{i,\up}^\dagger c_{i,\down}^\dagger|0\rangle$. It is clear that the fermion parity odd basis carries a projective representation of time reversal symmetry $T^2=-1$ while the fermion parity even basis carries a linear representation $T^2=1$. As a result, the time reversal symmetry group for interacting fermion systems has been an extension of the $\mathbb{Z}_2$ fermion parity symmetry group $\{I,P_f\}$, and the total symmetry group should consist of four group elements $\{I,T,T^2,T^3\}$ with $T^4=1$, which is a $\mathbb{Z}_4$ group. We note that the $\mathbb{Z}_2$ fermion parity symmetry can not be broken in \emph{local} interacting fermion systems, hence such a group extension can not be avoided. Since $1$D SPT phases are classified by the projective representation of the corresponding symmetry group\cite{XieSPT1,XieSPT2},
the Majorana spinons $(\gamma_\up,\gamma_\down)$ and $(\gamma_\up^\prime,\gamma_\down^\prime)$ on both ends must carry the projective representation of the bulk $\mathbb{Z}_4$ antiunitary symmetry with $T^4=1$, which leads to the $T^4=-1$ representation.

Possible experimental realizations of such an interesting TSC have been proposed by several groups recently\cite{1DTSC1,1DTSC2,1DTSC3,1DTSC4}.
In the following, we will show how to write down an explicit time reversal operator to realize the fractionalized $T^4=-1$ symmetry for a Majorana spinon.

\subsection{$T^4=-1$ time reversal symmetry}
For the pair of topological Majorana zero modes $\gamma_{\uparrow}$ and $\gamma_{\downarrow}$ on the left end, let us define the anti-unitary operator $T$ by $T=UK$, where $U$ is a unitary operator:
\begin{eqnarray}
U=\frac{1}{\sqrt{2}}(1+\gamma_\up \gamma_\down)=e^{\frac{\pi}{4}\gamma_\up\gamma_\down}\label{defTL}
\end{eqnarray}
Since $(\gamma_\up \gamma_\down)^\dagger= \gamma_\down\gamma_\up=-\gamma_\up \gamma_\down$, we have:
\begin{eqnarray}
U^\dagger=\frac{1}{\sqrt{2}}(1-\gamma_\up \gamma_\down)=e^{-\frac{\pi}{4}\gamma_\up\gamma_\down}
\end{eqnarray}
It is straightforward to verify that $U$ is a unitary operator:
\begin{eqnarray}
UU^\dagger=\frac{1}{2}(1+\gamma_\up \gamma_\down)(1-\gamma_\up \gamma_\down)=1
\end{eqnarray}
Furthermore, this new definition of time reversal operator gives rise to the correct transformation law for $\gamma_\up$ and $\gamma_{\down}$:
\begin{eqnarray}
T \gamma_{\uparrow} T^{-1}&=&
\frac{1}{2}(1+\gamma_\up \gamma_\down) \gamma_{\uparrow}(1-\gamma_\up \gamma_\down)=-\gamma_{\downarrow} \nonumber\\
T \gamma_{\downarrow} T^{-1}&=& \frac{1}{2}(1+\gamma_\up \gamma_\down) \gamma_{\downarrow}(1-\gamma_\up \gamma_\down)=\gamma_{\uparrow},
\end{eqnarray}
However, we notice that $T^2=\gamma_\up\gamma_\down\neq -1$ and satisfies:
\begin{eqnarray}
T^4=(\gamma_\up\gamma_\down)^2=-1
\end{eqnarray}
We call the two topological Majorana modes that carry the above $T^4=-1$ representation Majorana doublets, which can be viewed as a square root representation of the usual Kramers doublets. With such a definition of time reversal symmetry operator for a pair of topological Majorana modes, the symmetry protected nature becomes manifested, since a $T^4=-1$ projective representation can not be destroyed by time reversal preserving \emph{local} interactions. 

Similarly, for the pair of topological Majorana zero modes $\gamma_{\uparrow}^\prime,\gamma_{\downarrow}^\prime$ on the right end, $T$ can be defined by $T=U^\prime K$ with:
\begin{eqnarray}
U^\prime=\frac{1}{\sqrt{2}}(1+\gamma_\up^\prime \gamma_\down^\prime)=e^{\frac{\pi}{4}\gamma_\up^\prime\gamma_\down^\prime}\label{defTR}
\end{eqnarray}

The above definition of $T^4=-1$ time reversal operators on both ends can be applied to any physical site $i$ which contains two Majorana spinons
$(\gamma_{i,\up},\gamma_{i,\down})$ and $(\gamma_{i,\up}^\prime,\gamma_{i,\down}^\prime)$.
The total time reversal action is defined by $T=U_i\otimes U_i^\prime K$ with $U_i=e^{\frac{\pi}{4}\gamma_{i,\up}\gamma_{i,\down}}$ and $U_i^\prime=e^{\frac{\pi}{4}\gamma_{i,\up}^\prime\gamma_{i\down}^\prime}$. We have:
\begin{eqnarray}
T^2=\gamma_{i,\up}\gamma_{i,\down}\gamma_{i,\up}^\prime\gamma_{i,\down}^\prime=P^f_i=P^f_{i,L} P^f_{i,R}
\end{eqnarray}
with
\begin{eqnarray}
P^f_{i,L}=-i\gamma_{i,\up}\gamma_{i,\down};\quad P^f_{i,R}=i \gamma_{i,\up}^\prime\gamma_{i,\down}^\prime,
\end{eqnarray}
Here $P^f$ is the total fermion parity for a single physical site and $P^f_L$($P^f_R$) is fermion parity operator for the left(right) pair of Majorana spinon. The above definition of time reversal symmetry operator satisfies the requirement of $T^2=-1$ for fermion parity odd states while it satisfies $T^2=1$ for fermion parity even states.

\subsection{Representation theory of the $T^4=-1$ time reversal symmetry}
In the above, we use an algebraic way to construct the $T^4=-1$ symmetry, which will be very helpful for us to understand the underlying physics and provide us a simple way to do calculations. Now let us work out the explicit representation theory for the $T^4=-1$ time reversal symmetry.
We note that the two pairs of Majorana spinons on both ends allow us to define two complex fermions $c_L$ and $c_R$:
\begin{eqnarray}
c_L=\frac{1}{2}(\gamma_\up+i\gamma_\down);\quad c_R=\frac{1}{2}(\gamma_\up^\prime-i\gamma_\down^\prime)\label{cLR}
\end{eqnarray}
where $c_{L(R)}$ transforms nontrivially under the $T^4=-1$ symmetry. We have:
\begin{eqnarray}
T c_{L} T^{-1}&=&- i c_{L}^\dagger;\quad T c_{R} T^{-1}= i c_{R}^\dagger\nonumber\\ T c_{L}^\dagger T^{-1}&=& i c_{L};\quad T c_{R}^\dagger T^{-1}=- i c_{R}\label{TLR}
\end{eqnarray}
Since the $T$ operator only involves two Majorana operators, we are able to construct a precise two dimensional representation theory for the $T^4=-1$ symmetry. On the other hand, a projective representation can not be one dimensional, hence we must have:
\begin{eqnarray}
T|\t 0\rangle=UK|\t 0\rangle=U|\t 0\rangle=|\t 1\rangle\equiv c_{L(R)}^\dagger |\t 0\rangle \label{Rep1}
\end{eqnarray}
where $|\t 0\rangle$ is the vacuum of $c_{L(R)}$ fermion satisfying $c_{L(R)}|\t 0\rangle=0$ and $|\t1\rangle\equiv c_{L(R)}^\dagger |\t 0\rangle$. We also assume that the global phase of $|\t 0\rangle$ is fixed in such a way that the complex conjugate $K$ has a trivial action on it. From the relation Eq.(\ref{TLR}), it is straightforward to derive:
\begin{eqnarray}
T |\t 1\rangle &=&UK c_{L(R)}^\dagger |\t 0\rangle=U c_{L(R)}^\dagger |\t 0\rangle\nonumber\\
&=&T c_{L(R)}^\dagger T^{-1} T |\t 0\rangle=\pm ic_{L(R)}c_{L(R)}^\dagger |\t 0\rangle=\pm i|\t 0\rangle \label{Rep2}
\end{eqnarray}
Here the $+$ sign corresponds to $c_L$ and the $-$ sign corresponds to $c_R$.
Thus, on the basis of $|\t 0\rangle$ and $|\t 1\rangle$, we can derive the representation theory $T=UK$ with:
\begin{eqnarray}
U=\left(
    \begin{array}{cc}
      0 & 1 \\
      \pm i & 0 \\
    \end{array}
  \right),\label{RepT}
\end{eqnarray}
Clearly, the above representation satisfies $T^4=-1$.

\section{Topological Majorana zero modes in higher dimensions}
\label{sec: emergent spin}
In the previous section, we have discussed a simple example of a $1$D $T^2=-1$ TSC with topological Majorana zero modes on its ends. Topological Majorana zero modes exist in $\rm{DIII}$ class TSC in higher dimensions as well. In $2$D, it is well known that a single topological Majorana zero mode can emerge on the vortex core of a $p+ip$ or $p-ip$ TSC\cite{ReadMajorana}, but the time reversal symmetry is broken in this class of \emph{chiral} TSCs. Nevertheless, the $\rm{DIII}$ class TSC in $2$D consisting of a composition of a $p+ip$ and a $p-ip$ TSC with opposite spins\cite{2DTSC} can preserve the $T^2=-1$ time reversal symmetry. Apparently, the vortex core of such a TSC has a pair of topological Majorana zero modes $\gamma_\up$ and $\gamma_\down$ with opposite spins. Hence, we argue that they also carry a $T^4=-1$ representation of time reversal symmetry. As having been discussed in Ref. \cite{2DTSC}, a time reversal action on a single vortex core will change the \emph{local} fermion parity of the complex fermion zero mode $c_L=\gamma_\up+i\gamma_\down$ for the ground state wavefunction, therefore we expect the same representation theory Eq.(\ref{Rep1}),Eq.(\ref{Rep2}) and Eq.(\ref{RepT}) for the zero modes inside the vortex core, which satisfies $T^4=-1$. For the anti-vortex core with topological Majorana modes
$\gamma_\up^\prime$ and $\gamma_\down^\prime$, we can define a complex fermion mode $c_R=\gamma_\up-i\gamma_\down$ and derive the $T^4=-1$ representation theory as well. Now we see that the $c_L$/$c_R$ complex fermion is similar to the two complex fermion modes defined on the left/right end of the $1$D $T^2=-1$ TSC. The $T^4=-1$ time reversal operators for the Majorana spinons $(\gamma_\up,\gamma_\down)$ and $(\gamma_\up^\prime,\gamma_\down^\prime)$ can be defined by Eq.(\ref{defTL}) and Eq.(\ref{defTR}).

The $3$D analog of the vortex would be a hedgehog and the possibility of the emergence of a topological Majorana zero mode on the hedgehog has been proposed recently\cite{KaneMajorana}. However, there is an important difference in $3$D. Since the classical configuration of a hedgehog will have a divergent energy, the only way to introduce a UV cutoff is to couple the system to a gauge field, e.g., an $SU(2)$ gauge field\cite{3DTSC}. By turning on the $SU(2)$ gauge field, a single topological Majorana mode is not allowed\cite{Wittenanormaly} and we can only create a pair of topological Majorana zero modes. Therefore, the topological Majorana zero modes are unstable in the absence of time reversal symmetry(a mass term can be dynamically generated) and the analog of $p+ip$ TSC does not exist in $3$D. However, in the presence of $T^2=-1$ time reversal symmetry, the pair of topological Majorana zero modes $\gamma_\up$ and $\gamma_\down$ on the hedgehog can be stabilized(similar to the $1$D and $2$D cases, the mass term is forbidden by the time reversal symmetry) and we argue that they also carry a $T^4=-1$ time reversal symmetry according to the same reason as in $2$D -- the time reversal action changes the local fermion parity of the complex fermion mode $c_L=\gamma_\up+i\gamma_\down$ for the ground state wavefunction. The $\rm{DIII}$ class TSC in $3$D labeled by odd integers \cite{Kitaevperiod,Ryuperiod} could be a good candidate to realize a pair of topological Majorana zero modes on its hedgehog/anti-hedgehog. Detailed discussions of these interesting $3$D models are beyond the scope of this paper and will be presented elsewhere.

Finally, we would like to point out an important difference for the topological Majorana zero modes between $1$D and higher dimensions. In $1$D, for a generic Hamiltonian, the zero modes are only well defined in the infinite long chain limit. However, in $2(3)$D, the distance between vortex(hedgehog) and anti-vortex(anti-hedgehog) can be finite(but much larger than penetration depth) since the zero modes are well defined bound states and they can be regarded as \emph{local} particles.

\begin{figure}[tb]
\begin{center}
\vskip -0.5cm
\includegraphics[width=7.5cm]{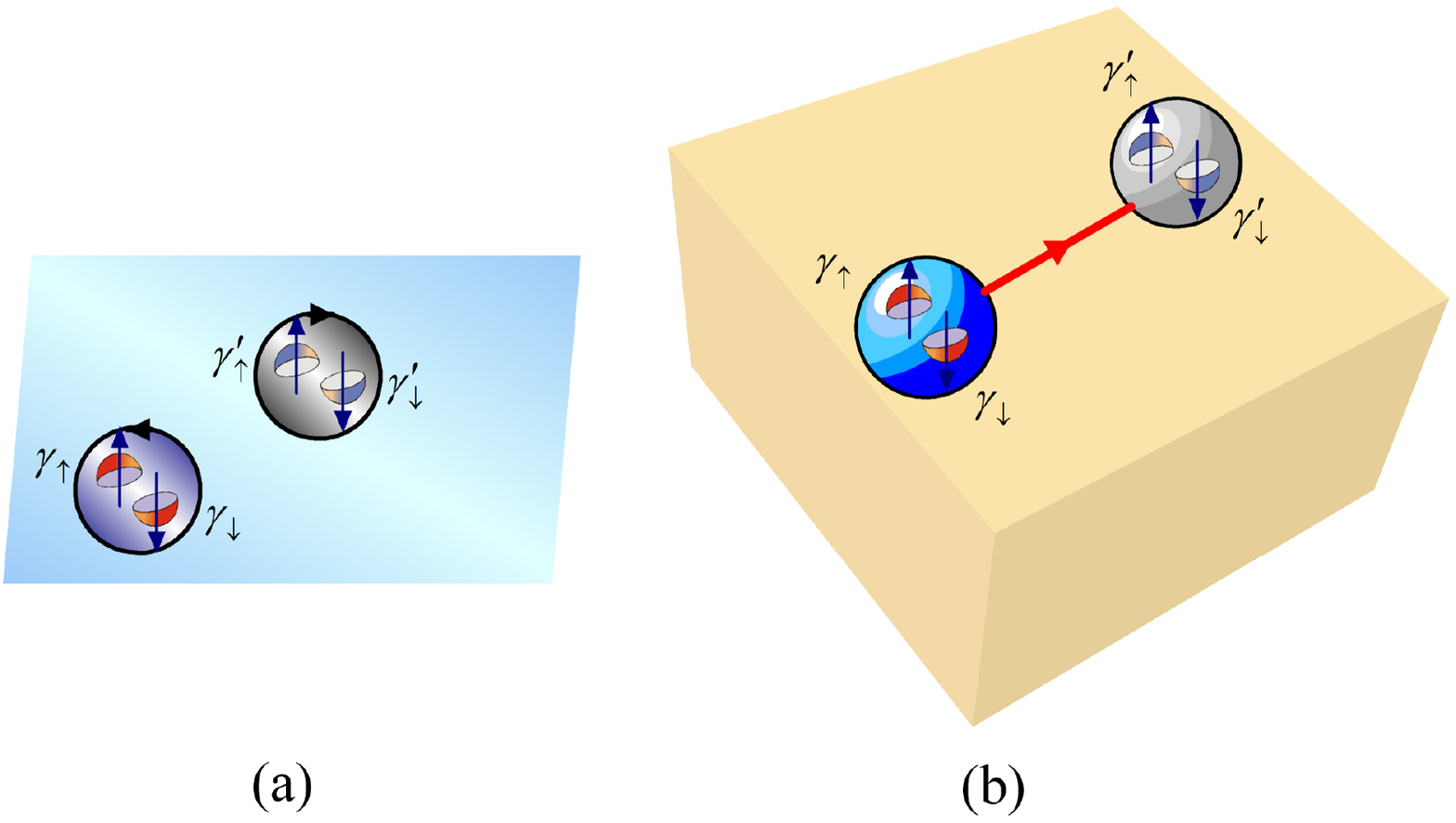}
\caption{(color online)Topological Majorana zero modes in $2$D and $3$D can be realized as the bound states on the vortex/anti-vortex core and hedgehog/anti-hedgehog core of $\rm{DIII}$ class TSC. The red line in (b) represents a quantized flux line that connects a pair of hedgehog and anti-hedgehog.
}\label{hihgerD}
\end{center}
\end{figure}

\section{Emergent relativistic dispersion, $SU(2)$ spin at quantum criticality}
\label{appemergent}
\subsection{Deconfined topological Majorana zero modes in $1$D}
\begin{figure}[bt]
\begin{center}
\includegraphics[width=7.5cm]{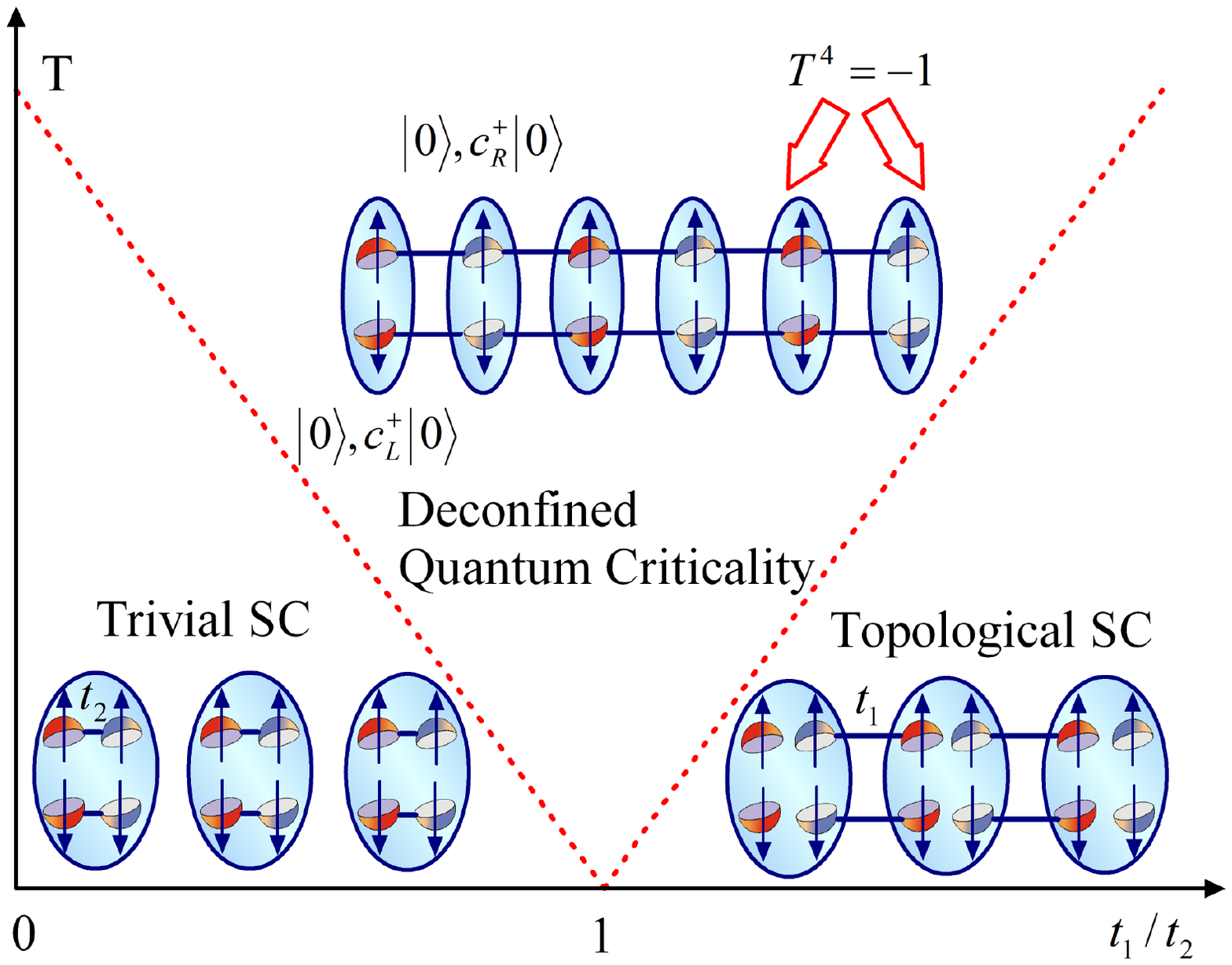}
\caption{(color online) At the deconfined quantum critical point, the proliferation of topological Majorana zero modes leads to the emergence of relativistic dispersion and $SU(2)$ (pseudo) spin rotational symmetry.
}\label{deconfine}
\end{center}
\end{figure}
In the following we will show that relativistic dispersion and $SU(2)$ spin rotational symmetry will emerge at a quantum critical point where topological Majorana zero modes are proliferated.
Let us begin with a $1$D model by adding a chemical potential term to the Hamiltonian Eq.(\ref{H}):
\begin{eqnarray}
H^\prime&=&\sum_{i=1}^N\sum_{\sigma} \left(c_{i,\sigma}-c_{i,\sigma}^\dagger \right)\left(c_{i+1,\sigma}+c_{i+1,\sigma}^\dagger \right)\nonumber\\&-&2\mu \sum_{i=1}^N\sum_{\sigma}(c_{i,\sigma}^\dagger c_{i,\sigma}-\frac{1}{2}),
\end{eqnarray}
As having been discussed in Ref. \cite{KitaevMajorana}, a phase transition occurs at $\mu=1$ and the system becomes a trivial superconductor when
$\mu>1$. In terms of Majorana operators, we will have a simple picture to visualize the above phase transition.
\begin{eqnarray}
H^\prime=\sum_{i=1}^N\sum_{\sigma} i\sigma\gamma_{i,\sigma}^{\prime}\gamma_{i+1,\sigma}+\mu\sum_{i=1}^N\sum_{\sigma} i\sigma\gamma_{i,\sigma}^{\prime}\gamma_{i,\sigma},
\end{eqnarray}

As seen in Fig. \ref{deconfine}, in the limit where the on site hopping $t_2\equiv\mu$ is dominant, the above Hamiltonian describes a trivial superconductor, while in the limit where the inter site hopping $t_1(=1)$ is dominant, it describes a topological superconductor with topological Majorana modes confined on both ends. At the phase transition point $t_2=t_1=1$, the Majorana spinon with $T^4=-1$ time reversal symmetry becomes deconfined.
The analog of such a deconfined quantum critical phenomenon has been known for a long time in $1$D spin chain models with $T^2=1$ time reversal symmetry, e.g., in certain spin-$1$ chain systems\cite{Haldanecritical}, $1/2$ spinon with $T^2=-1$  time reversal symmetry on its ends becomes deconfined at the phase transition point.

At low energy, the critical theory has emergent relativistic dispersion and $SU(2)$ (pseudo) spin. In terms of $c_{L(R)}$ fermions, we can rewrite the critical Hamiltonian as:

\begin{eqnarray}
H_{1D}&=&i\sum_{ i}(c_{i,L}^\dagger c_{i+1,R}-c_{i+1,R}^\dagger c_{i,L})\nonumber\\&+&i\sum_{ i}(c_{i,L}^\dagger c_{i,R}-c_{i,R}^\dagger c_{i,L})
\end{eqnarray}
In momentum space, the above Hamiltonian can be diagonalized by:
\begin{widetext}
\begin{eqnarray}
H_{1D}=\sum_k (c_L^\dagger(k),c_R^\dagger(k))\left(
                                            \begin{array}{cc}
                                              0 & i(1+e^{ik}) \\
                                              -i(1+e^{-ik}) & 0 \\
                                            \end{array}
                                          \right)\left(\begin{array}{c}
                                                         c_L(k) \\
                                                         c_R(k)
                                                       \end{array}\right)
\end{eqnarray}
\end{widetext}
 The above Hamiltonian has one positive energy mode and one negative energy mode with $E_k=\pm 2t|\cos \frac{k}{2}|$. The dispersion relation is relativistic around the momentum points $k=\pm \pi$. The particle and hole excitations in such a system form an $SU(2)$ doublet. However, for a 1D chain, the $SU(2)$ (pseudo) spin rotational symmetry does not carry angular momentum and is a purely internal symmetry. However, in 3D, the deconfinement of topological Majorana zero modes might lead to the emergence of $SU(2)$ spin carrying angular momentum, which is no longer a purely internal symmetry. Since the universal conformal field theory description for deconfinement topological Majorana zero modes in 3D is still unknown so far, let us just construct a particular model here to illustrate the main idea.

\subsection{Deconfined topological Majorana zero modes in $3$D and emergent relativistic dispersion, $SU(2)$ spin}
\begin{figure}[tb]
\begin{center}
\includegraphics[width=8cm]{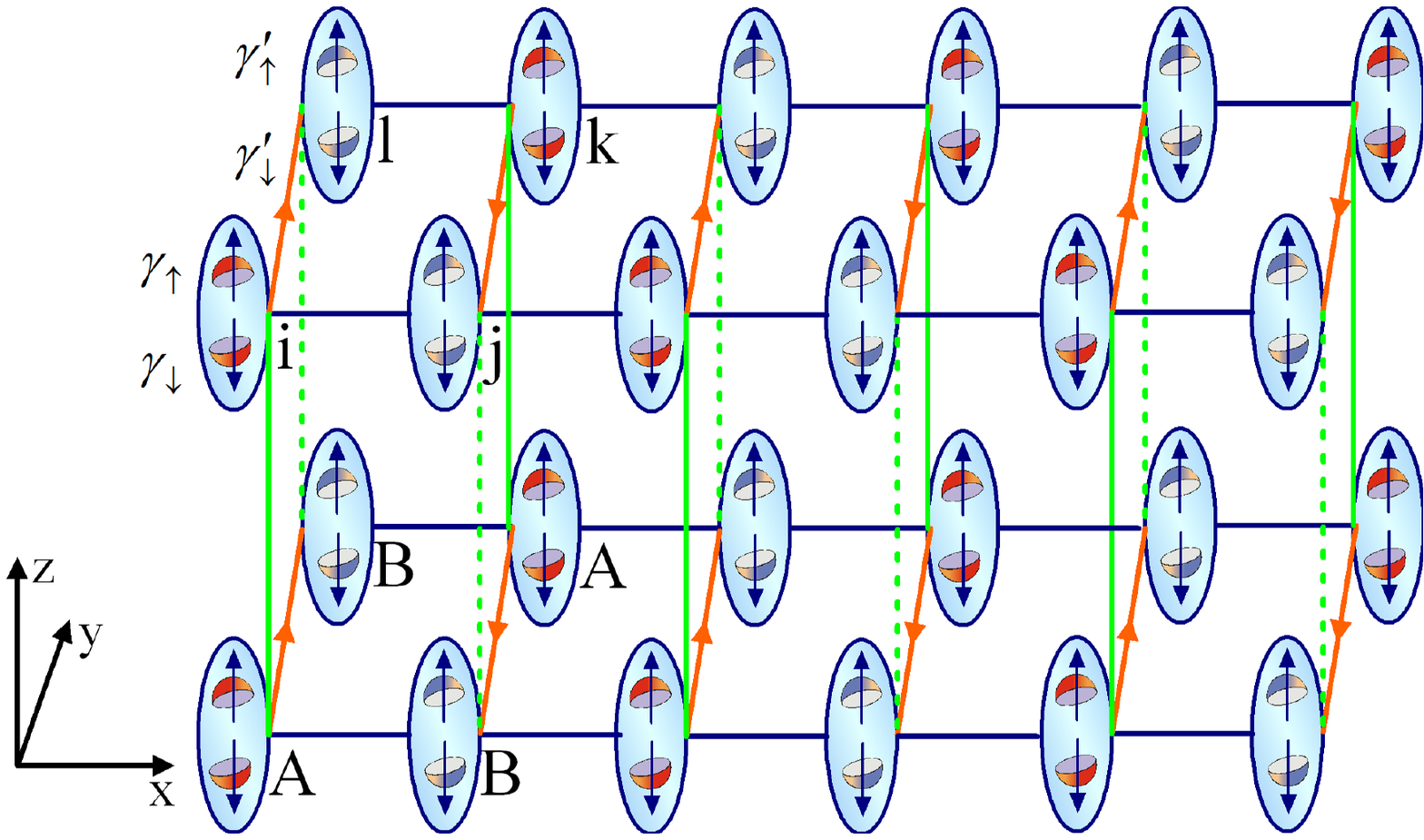}
\caption{(color online)A $3$D hedgehog/anti-hedgehog cubic lattice. Red dots represent the pair of topological Majorana modes $(\gamma_\up, \gamma_\down)$ on the hedgehog and blue dots represent the pair of topological Majorana modes $(\gamma_\up^\prime, \gamma_\down^\prime)$ on the anti-hedgehog. Solid/dashed lines represent the hopping amplitude $1/-1$. Lines with arrows represent the hopping amplitudes $\pm i$. Multiplications of the hopping amplitudes surround a square surface give rise to $-1$, e.g., $t_{ij}t_{jk}t_{kl}t_{li}=-1$.
Such a hopping amplitudes pattern is the so called $\pi$-flux pattern.
}\label{3Dchain}
\end{center}
\end{figure}

First, we construct a $3$D cubic lattice model consisting of hedgehog occupied sublattice A and anti-hedgehog occupied sublattice B, as seen in Fig \ref{3Dchain}. We use red dots to represent the pair of topological Majorana modes $(\gamma_\up, \gamma_\down)$ on the hedgehog and blue dots to represent the pair of topological Majorana modes $(\gamma_\up^\prime, \gamma_\down^\prime)$ on the anti-hedgehog.
Similar to the $1$D cases, we then turn on the hopping among those topological Majorana modes and consider the following Hamiltonian:
\begin{widetext}
\begin{eqnarray}
H_{3D}&=&-\sum_{\v i\in A; \v j=\v i \pm \hat{\v x}}\left( i\gamma_{\v i,\up} \gamma_{\v j,\down}^\prime +i\gamma_{\v i,\down} \gamma_{\v j,\up}^\prime\right)+\sum_{\v i\in A; \v j=\v i \pm \hat{\v y}}\left( i\gamma_{\v i,\up} \gamma_{\v j,\up}^\prime -i\gamma_{\v i,\down} \gamma_{\v j,\down}^\prime\right)
\nonumber\\&+&\sum_{ \v i \in A; \v j=\v i+ \hat{\v z}}\left( i\gamma_{\v i,\up} \gamma_{\v j,\down} -i\gamma_{\v i,\down} \gamma_{\v j,\up}\right)+\sum_{ \v i \in B; \v j=\v i+ \hat{\v z}}\left(i\gamma_{\v i,\up}^\prime \gamma_{\v j,\down}^\prime  -i\gamma_{\v i,\down}^\prime  \gamma_{\v j,\up}^\prime \right)\label{3DMajorana}
\end{eqnarray}
In terms of complex fermions $c_{\v i, L}=\gamma_{\v i, \up}+i\gamma_{\v i,\down}$ and $c_{\v i, R}=\gamma_{\v i,\up}^\prime-i\gamma_{\v i\down}^\prime$, we have:
\begin{eqnarray}
H_{3D}&=&\sum_{\v i\in A; \v j=\v i \pm \hat{\v x}}\left( c_{L,\v i}^\dagger c_{R,\v j}+c_{R,\v j}^\dagger c_{L,\v i}\right)+i\sum_{\v i\in A; \v j=\v i \pm \hat{\v y}}\left( c_{L,\v i}^\dagger c_{R,\v j}-c_{R,\v j}^\dagger c_{L,\v i}\right)
\nonumber\\&+&\sum_{ \v i \in A; \v j=\v i+ \hat{\v z}}\left( c_{L,\v i}^\dagger c_{L,\v j}+ c_{L,\v j}^\dagger c_{L,\v i}\right)-\sum_{ \v i \in B; \v j=\v i+ \hat{\v z}}\left(c_{R,\v i}^\dagger c_{R,\v j}+c_{R,\v j}^\dagger c_{R,\v i}\right)\label{3Dlattice}
\end{eqnarray}

The special hopping pattern in the above Hamiltonian is one way to realize the so called $\pi$-flux pattern, namely, a pattern with the enclosed flux $\pi$ on each face of the cubic lattice.
The Hamiltonian is invariant under the time reversal symmetry $\t T=T^{(-)^{i_z}}$. Without such a twisted definition of the time reversal symmetry, the fermion hopping in the $z$-direction will change sign under the time reversal. It is clear that such a twisted definition is allowed because we can choose either $T$ or $T^{-1}$ as the definition of the time reversal symmetry.

In momentum space, we have:
\begin{eqnarray}
H_{3D}=\sum_{\v k} (c_L^\dagger(\v k),c_R^\dagger(\v k))\left(
                                            \begin{array}{cc}
                                              2\cos k_z & 2\cos k_x+2i\cos k_y \\
                                              2\cos k_x-2i\cos k_y & -2\cos k_z  \\
                                            \end{array}
                                          \right)\left(\begin{array}{c}
                                                         c_L(\v k) \\
                                                         c_R(\v k)
                                                       \end{array}\right)
\end{eqnarray}
The above Hamiltonian has one positive energy mode and one negative energy mode with:
\begin{eqnarray}
E_{\v k}=\pm 2\sqrt{\cos^2 k_x+\cos^2 k_y+\cos^2 k_z},
\end{eqnarray}
Around the momentum point $\v k_0=(\pi/2,\pi/2,\pi/2)$, the above Hamiltonian describes a chiral Weyl fermion:
\begin{eqnarray}
H^{eff}_{(\pi/2,\pi/2,\pi/2)}=2\sum_{\v k} (c_L^\dagger(\v k),c_R^\dagger(\v k))\left(
                                            \begin{array}{cc}
                                              \bar k_z & \bar k_x+i\bar k_y \\
                                             \bar k_x-i\bar k_y & -\bar k_z  \\
                                            \end{array}
                                          \right)\left(\begin{array}{c}
                                                         c_L(\v k) \\
                                                         c_R(\v k)
                                                       \end{array}\right)
\end{eqnarray}
where $\v k=\v k_0+ \bar {\v k}$. It is clear that the above Hamiltonian has a relativistic dispersion $E_{\v k}=\pm 2|\bar {\v k}|$ and an emergent $SU(2)$ spin carrying angular momentum.
\end{widetext}

In the above, we construct a particular $3$D hedgehog/anti-hedgehog lattice model with proliferated topological Majorana zero modes. Other models with deconfined topological Majorana modes have also been considered recently, e.g., the fermion dimer model\cite{gapless3DTSC} and the Majorana flat band model in certain gapless TSC\cite{gaplessMajorana}. However, one of the most important features in our model is that it has a sublattice structure, and the sublattice degeneracy naturally leads to an $SU(2)$ spin degree of freedom at low energy. Actually, our model can be viewed as the $3$D analog of the $2$D graphene system, where the valley degeneracy becomes the emergent $SU(2)$ spin at low energy. But why hedgehog/anti-hedgehog lattice models with a sublattice structure are more natural than those models without a sublattice structure? One possible reason is that the hedgehog and anti-hedgehog pair are always confined in a superconductor\cite{monopoleconfinement}, therefore any stable $3$D hedgehog/anti-hedgehog lattice model must contain a hedgehog and anti-hedgehog pair per unit cell.

Our analysis for condensed matter systems implies that the presence of $SU(2)$ spin at low energy has a deep relationship with the sublattice structure at cutoff scale. A very interesting question is whether the $SU(2)$ spin for all the fundamental particles arises from a similar discrete structure at cutoff scale. Unfortunately, SM with an explicit cutoff is absent so far due to the chiral fermion problem. Recent development on the vanishing of non-perturbative anomaly in chiral $SO(10)$ gauge theory\cite{wenchiral1,wenchiral2} sheds new light on this long standing hard problem and makes it possible to realize chiral gauge theory in lattice models with strong fermion interactions.

Finally, although lattice models could be thought as a natural venue to regulate a quantum field theory, any pre-assumed lattice structure for space-time will break the Lorentz invariance. To overcome this difficulty, the topological non-linear sigma model will be a promising candidate. Important progress along this direction has been made recently\cite{XieSPT3,XieSPT4}, even with fermions\cite{Gusuper}.

\section{Fractionalized parity and charge conjugation symmetry}
\label{sec: superCPT}
So far, we have constructed concrete condensed matter models with topological Majorana zero modes carrying $T^4=-1$ time reversal symmetry on point-like defects in certain classes of TSCs. In last section, we have also shown that the proliferation of topological Majorana zero modes will lead to a chiral Weyl fermion with emergent relativistic dispersion and $SU(2)$ spin at low energy. Since neutrinos are described as the chiral Weyl fermion, it is very natural to ask if they can be interpreted as the (proliferated) topological Majorana zero modes. However, such a conjecture could be very challenging as it requires a strongly correlated vacuum instead of the trivial vacuum assumed in traditional quantum field theory. Nevertheless, in the semiclassical limit, it is still possible to investigate other fractionalized (discrete) symmetries carried by topological Majorana zero modes and discuss the interesting physical consequence. In this section, we limit our discussion at the single particle level, and the generalization into the quantum field theory will be presented in the next section.

As having been discussed in last section, the confinement of hedgehog and anti-hedgehog pair in $3$D superconductor suggests that the four topological Majorana zero modes $\gamma_\up,\gamma_\up^\prime,\gamma_\down$ and $\gamma_\down^\prime$ identify the local degrees of freedom with respect to translational symmetry.(For a lattice model, those are the degrees of freedom in a unit cell.) On the other hand, a relativistic Majorana fermion is a four component Lorentz spinon; hence, it is natural to investigate the full symmetry properties of the four dimensional zero energy subspace expanded by the four topological Majorana zero modes $\gamma_\up,\gamma_\up^\prime,\gamma_\down$ and $\gamma_\down^\prime$. Particularly, we will discuss the other two fundamental discrete symmetries -- parity and charge conjugation.

\subsection{$P^4=-1$ parity symmetry for a pair of topological Majorana zero modes}
For a single particle, we only consider the parity symmetry as a $\mathbb{Z}_2$ action on the internal degrees of freedom, and in quantum field theory, we will include its action on coordinates as well.
Interestingly, in the zero energy subspace expanded by four topological Majorana zero modes, we can define a $P^4=-1$ symmetry for each parity pair of topological Majorana zero modes $\gamma_\up,\gamma_\up^\prime$ or $\gamma_\down,\gamma_\down^\prime$. The reason why we can have such a fractionalized parity symmetry for topological Majorana zero modes is the same as the reason for time reversal symmetry. The parity symmetry for an interacting spin-$1/2$ fermion system is actually $P^2=P^f$. Therefore, for the Fock basis $c_\up^\dagger |0\rangle,c_\down^\dagger|0\rangle$ and $|0\rangle,c_\up^\dagger c_\down^\dagger|0\rangle$, the parity odd sector satisfies $P^2=-1$ while the parity even sector satisfies $P^2=1$. Here the complex fermion operators $c_\up $ and $c_\down$ are defined by:
\begin{eqnarray}
c_\up=\gamma_\up+i\gamma_\up^\prime;\quad c_\down=\gamma_\down-i\gamma_\down^\prime,
\end{eqnarray}
which gives rise to a natural notion of spin basis out of four topological Majorana zero modes.

The explicit construction of $P^4=-1$ operator for a pair of topological Majorana zero modes is very similar to that for the $T^4=-1$ time reversal symmetry. For the pair of topological Majorana modes $\gamma_\up,\gamma_\up^\prime$ and $\gamma_\down,\gamma_\down^\prime$, their parity operators are defined by:
\begin{eqnarray}
P_{\up\up^\prime}&=&\frac{1}{\sqrt{2}}(1+\gamma_\up\gamma_\up^\prime )=e^{\frac{\pi}{4}\gamma_\up\gamma_\up^\prime}; \nonumber\\ P_{\down \down^\prime}&=&\frac{1}{\sqrt{2}}(1-\gamma_\down \gamma_\down^\prime)=e^{-\frac{\pi}{4}\gamma_\down\gamma_\down^\prime},
\end{eqnarray}
We see such a definition satisfies $P_{\up\up^\prime(\down \down^\prime)}^4=-1$ for each pair of topological Majorana modes. The total parity action on the four topological Majorana zero modes is defined by $P=P_{\up\up^\prime}\otimes P_{\down \down^\prime}$. Its action on the four topological Majorana modes reads:
\begin{eqnarray}
P \gamma_{\uparrow} P^{-1}&=&
-\gamma_{\uparrow}^\prime; \quad
P \gamma_{\downarrow} P^{-1}= \gamma_{\downarrow}^\prime\nonumber\\
P \gamma_{\uparrow}^\prime P^{-1}&=&
\gamma_{\uparrow}; \quad
P \gamma_{\downarrow}^\prime P^{-1}= -\gamma_{\downarrow},
\end{eqnarray}
It is easy to verify that the complex fermions $c_\up $ and $c_\down$ representing the spin basis transform in an expected way:
\begin{eqnarray}
P c_{\uparrow} P^{-1}&=& i
c_{\uparrow}; \quad
P c_{\downarrow} P^{-1}= i c_{\downarrow}\nonumber\\
 P c_{\uparrow}^\dagger P^{-1}&=&
-i c_{\uparrow}^\dagger; \quad
P c_{\downarrow}^\dagger P^{-1}= -i c_{\downarrow}^\dagger,
\end{eqnarray}
We note that although the spin of a particle does not change under parity, there could be a nontrivial phase factor for the spin-$1/2$ particle.
On the other hand, $c_L=\gamma_\up+i\gamma_\down$ and $c_R=\gamma_\up^\prime-i\gamma_\down^\prime$ transform like a neutrino and an anti-neutrino pair:
\begin{eqnarray}
P c_L P^{-1}&=&
-c_R; \quad
P c_R P^{-1}= c_L\nonumber\\
P c_L^\dagger P^{-1}&=&
-c_R^\dagger; \quad
P c_R^\dagger P^{-1}= c_L^\dagger\label{P}
\end{eqnarray}
Our definition of parity operator is comparable with the time reversal operator $PTP^{-1}=P^f T$ with $T=e^{\frac{\pi}{4}\gamma_\up\gamma_\down}e^{\frac{\pi}{4}\gamma_\up^\prime\gamma_\down^\prime}K$, and $P^f=\gamma_\up\gamma_\down \gamma_\up^\prime\gamma_\down^\prime$ is the total fermion parity operator.

\subsection{$\overline C^4=-1$ charge conjugation symmetry for a pair of topological Majorana zero modes}
Since the Majorana fermion describes a neutral particle, the charge conjugation action is trivial from a traditional perspective.
Strikingly, we find a way to define a nontrivial $\overline C^4=-1$ charge conjugation symmetry for a pair of topological Majorana zero modes.
Similar to the $T^4=-1$/$P^4=-1$ time reversal/parity symmetry, for each pair of topological Majorana zero modes with opposite spins, we can define a $\overline C^4=-1$ charge conjugation operator:
\begin{eqnarray}
\overline C_{\up\down^\prime}&=&\frac{1}{\sqrt{2}}(1+\gamma_\up \gamma_\down^\prime)=e^{\frac{\pi}{4}\gamma_\up\gamma_\down^\prime};\nonumber\\ \overline C_{\down\up^\prime}&=&\frac{1}{\sqrt{2}}(1+\gamma_\down \gamma_\up^\prime)=e^{\frac{\pi}{4}\gamma_\down\gamma_\up^\prime},
\end{eqnarray}
and the total action of charge conjugation symmetry on four topological Majorana zero modes is $\overline C=\overline C_{\up\down^\prime}\otimes \overline C_{\down\up^\prime}$.
It is straightforward to verify:
\begin{eqnarray}
\overline C \gamma_{\uparrow} \overline C^{-1} &=&
-\gamma_{\downarrow}^\prime; \quad
\overline C \gamma_{\downarrow} \overline C^{-1}= -\gamma_{\uparrow}^\prime\nonumber\\
\overline C \gamma_{\uparrow}^\prime \overline C^{-1} &=&
\gamma_{\downarrow}; \quad
\overline C \gamma_{\downarrow}^\prime \overline C^{-1}= \gamma_{\uparrow},
\end{eqnarray}
which implies:
\begin{eqnarray}
\overline C c_{\uparrow} \overline C^{-1} &=&
ic_{\downarrow}^\dagger; \quad
\overline C c_{\downarrow} \overline C^{-1}= -ic_{\uparrow}^\dagger\nonumber\\
\overline C c_{\uparrow}^\dagger \overline C^{-1} &=&
-i c_{\downarrow}; \quad
\overline C c_{\downarrow}^\dagger \overline C^{-1}=ic_{\uparrow},
\end{eqnarray}
and
\begin{eqnarray}
\overline C c_{L} \overline C^{-1} &=&
-ic_{R}; \quad
\overline C c_{R} \overline C^{-1}= -i c_{L}\nonumber\\
\overline C c_{L}^\dagger \overline C^{-1} &=&
i c_{R}^\dagger; \quad
\overline C c_{R}^\dagger \overline C^{-1}=i c_{L}^\dagger,\label{C}
\end{eqnarray}
We note that for the spin basis $c_{\up(\down)}$, the charge conjugation acts as a particle-hole symmetry;
however, for the $c_{L(R)}$ basis it acts like a neutrino and anti-neutrino exchange symmetry(if we interpret $c_{L}$ as a neutrino and $c_{R}$ as an anti-neutrino). Similar to the commutation relation between time reversal and parity symmetry, the $\overline C^4=-1$ charge conjugation symmetry also commutes with the other two symmetries up to a total fermion parity.
\begin{eqnarray}
\overline CT \overline C^{-1}=P^fT;\quad \overline CP \overline C^{-1}=P^f P
\end{eqnarray}

\subsection{$\overline C,P,T$ super algebra for a Majorana fermion}
Let us summarize the closed algebraic relation of $\overline C,P,T,$ and $P^f$ symmetries for a Majorana fermion formed by four topological Majorana zero modes.
\begin{eqnarray}
\overline C^2&=&P^f;\quad P^2=P^f ;\quad T^2=P^f;\quad {(P^f)}^2=1\nonumber\\
TP^f&=&P^fT;\quad PP^f=P^fP;\quad \overline CP^f=P^f \overline C\nonumber\\
TP &=& P^fPT;\quad T\overline C=P^f\overline CT;\quad P\overline C=P^f \overline CP, \label{sual}
\end{eqnarray}
The above algebra satisfied by the $\overline C,P,T$ symmetries is indeed a super algebra, which can be regarded as a super extension of the usual charge conjugation, parity and time reversal symmetries over the fermion parity symmetry $P^f$. This super algebra is one of the central results of this paper. It arises from the topological nature of the topological Majorana zero modes and reflects the strongly correlated nature of the vacuum.

In next section, we will show that the above $\overline C,P,T$ super algebra is also applicable for Majorana field. In quantum field theory, such a super extension is allowed because $P^f$ is not a physical observable, or in other words, there is no way to measure the total fermion parity of a quantum state since any physical process must preserve fermion parity symmetry. From a traditional point of view, our results suggest that the $\overline C,P,T$ transformations for Majorana field can be different from a Dirac field, just like a scalar field and a Dirac field have very different $C,P,T$ transformations. Therefore, a Majorana field with a topological origination has a completely new physical meaning and indicates a strongly correlated vacuum, despite the equivalence between Majorana representation and Weyl representation\cite{eq}.

In addition to the fundamental discrete symmetries $\overline C,P,T$, we can also define a spin rotational symmetry in the spin basis, where $c_\up^\dagger |0\rangle,c_\down^\dagger|0\rangle$ carry spin-$1/2$ while $|0\rangle,c_\up^\dagger c_\down^\dagger|0\rangle$ carry spin-0. Therefore, the $SU(2)$ spin operator $\mathbf{S}$ can be naturally defined by:
\begin{eqnarray}
S^\alpha=\frac{1}{2}\sum_{\sigma,\sigma^\prime}c_\sigma^\dagger \tau_{\sigma\sigma^\prime}^\alpha c_{\sigma^\prime}; \alpha=x,y,z
\end{eqnarray}
where $\tau^\alpha$ is the usual Pauli matrix.
It is easy to verify that:
\begin{eqnarray}
T \mathbf{S} T^{-1}=-\mathbf{S};\quad P \mathbf{S} P^{-1}=\mathbf{S};\quad  \overline C \mathbf{S} \overline C^{-1}=\mathbf{S},\label{comparable}
\end{eqnarray}
The above nice property makes the $\overline C,P,T$ symmetries commute with the $SU(2)$ spin rotational symmetry and allows us to generalize the $\overline CPT$ super algebra into the relativistic quantum field theory.

\section{$\overline C,P,T$ symmetries for Majorana field}
\label{sec: field theory}
\subsection{$\overline C,P,T$ symmetries for relativistic quantum field theory}
Let us implement the $\overline C,P,T$ symmetries to a Majorana field. We choose four real gamma matrices:
\begin{eqnarray}
\gamma_0&=&-i\rho_z \otimes \sigma_y;\quad \gamma_1=-I \otimes \sigma_x ;\nonumber\\  \gamma_2&=&\rho_y \otimes \sigma_y ;\quad\gamma_3=I \otimes \sigma_z,
\end{eqnarray}
where $\rho$ and $\sigma$ are Pauli matrices and $I$ is the identity matrix. We can define a real $\gamma_5$ by:
\begin{eqnarray}
\gamma_5=\gamma_0\gamma_1\gamma_2\gamma_3=i\rho_x\otimes\sigma_y
\end{eqnarray}
The four component Majorana field describing the pair of complex fermions $c_L$ and $c_R$ reads:
\begin{eqnarray}
\psi_c(x)= \left(
        \begin{array}{c}
         \xi(x)\\
        \eta(x) \\
        \end{array}
      \right),
\end{eqnarray}
where
\begin{eqnarray}
   \xi(x)= \left(
        \begin{array}{c}
          \gamma_\up(x)  \\
          \gamma_\down(x)  \\
        \end{array}
      \right);\quad
\eta(x)= \left(
        \begin{array}{c}
          -\gamma_\up^\prime(x)  \\
          \gamma_\down^\prime(x) \\
        \end{array}
      \right),
\end{eqnarray}
Here the Majorana spinon basis $\xi(x)$ and $\eta(x)$ are equivalent to complex fermions $c_L$ and $c_R$, which gives rise to a natural notion of neutrino and anti-neutrino.

The (equal time) canonical commutation relation reads:
\begin{eqnarray}
\{\psi^\dagger_c (\v x),\psi_c (\v y)\}=2\delta^{(3)}(\v x-\v y).
\end{eqnarray}
In terms of real Majorana modes $\gamma_{\sigma}(\v x)$ and $\gamma_{\sigma}^\prime(\v x)$, we have:
\begin{eqnarray}
\{\gamma_{\sigma}(\v x),\gamma_{\sigma^\prime}^\prime(\v y)\}&=&0; \nonumber\\ \{\gamma_{\sigma}(\v x),\gamma_{\sigma^\prime}(\v y)\}&=&2\delta^{(3)}(\v x-\v y)\delta_{\sigma\sigma^\prime},
\end{eqnarray}
which is the continuum version of the commutation relation Eq.(\ref{commutate}).
The $\overline C,P,T$ symmetry operators can be defined by:
\begin{eqnarray}
\overline C&=&\prod_{\v x}e^{\frac{\pi}{4}\gamma_{\up}(\v x)\gamma_{\down}^\prime(\v x)}e^{\frac{\pi}{4}\gamma_{\down}(\v x)\gamma_{\up}^\prime
(\v x)}\nonumber\\&=&e^{\frac{\pi}{4}\int d^3 x\gamma_{\up}(\v x)\gamma_{\down}^\prime(\v x)}e^{\frac{\pi}{4}\int d^3 x\gamma_{\down}(\v x)\gamma_{\up}^\prime
(\v x)}\nonumber\\
P&=&U_PP_0=\prod_{\v x}e^{\frac{\pi}{4}\gamma_{\up}(\v x)\gamma_{\up}^\prime(\v x)}e^{-\frac{\pi}{4}\gamma_{\down}(\v x)\gamma_{\down}^\prime
(\v x)}P_0\nonumber\\&=&e^{\frac{\pi}{4}\int d^3x \gamma_{\up}(\v x)\gamma_{\up}^\prime(\v x)}e^{-\frac{\pi}{4}\int d^3 x\gamma_{\down}(\v x)\gamma_{\down}^\prime
(\v x)}P_0\nonumber\\
T&=&U_TK=\prod_{\v x}e^{\frac{\pi}{4}\gamma_{\up}(\v x)\gamma_{\down}(\v x)}e^{\frac{\pi}{4}\gamma_{\up}^\prime(\v x)\gamma_{\down}^\prime
(\v x)}K\nonumber\\&=&e^{\frac{\pi}{4}\int d^3 x\gamma_{\up}(\v x)\gamma_{\down}(\v x)}e^{\frac{\pi}{4}\int d^3 x\gamma_{\up}^\prime(\v x)\gamma_{\down}^\prime
(\v x)}K\nonumber\\
P^f&=&\prod_{\v x}\gamma_{\up}(\v x)\gamma_{\down}(\v x)\gamma_\up^\prime(\v x)\gamma_\down^\prime(\v x)=\overline C^2=T^2=P^2. \label{CPT}
\end{eqnarray}
Here $P_0$ is the action on the spacial coordinates with $P_0 \v x P_0^{-1}=-\v x$. It is easy to check that the above $\overline C,P,T$ symmetry operators satisfy the super algebra Eq.(\ref{sual}).

The transformations of the Majorana field under the above $\overline C,P,T$ symmetries can also be derived:
\begin{eqnarray}
\overline C \psi_c(x) \overline C^{-1}&=&  \left(
        \begin{array}{c}
         -\epsilon\eta(x)\\
        -\epsilon\xi(x) \\
        \end{array}
      \right)
= -\gamma_5\psi_c( x);\nonumber\\ P
\psi_c(x) P^{-1}&=&  \left(
        \begin{array}{c}
         \eta(\t x)\\
         -\xi(\t x) \\
        \end{array}
      \right)=\gamma_0\gamma_5 \psi_c(\t x);\nonumber\\ T \psi_c(x) T^{-1}&=& \left(
        \begin{array}{c}
         -\epsilon\xi(-\t x)\\
         \epsilon\eta(-\t x) \\
        \end{array}
      \right)=\gamma_0 \psi_c(-\t x),\nonumber\\\label{CPT}
\end{eqnarray}
where $\t x=(t,-\v x)$. Let us consider the Majorana field Lagrangian in the massless limit:
\begin{eqnarray}
\mathcal{L}_0=\frac{1}{4}\overline\psi_c(x)i\gamma_\mu \partial_\mu \psi_c(x); \quad \overline\psi_c(x)=\psi^\dagger_c(x)\gamma_0, \label{free}
\end{eqnarray}
Apparently, $\mathcal{L}_0$ is invariant under the $\overline C,P,T$ symmetries:
\begin{eqnarray}
\overline  C \mathcal{L}_0(x) \overline C^{-1}&=&\mathcal{L}_0(x); \quad P \mathcal{L}_0(x) P^{-1}=\mathcal{L}_0(\t x);\nonumber\\ T \mathcal{L}_0(x) T^{-1}&=&\mathcal{L}_0(-\t x),
\end{eqnarray}

\subsection{Charge conjugation as a $\mathbb{Z}_2$ gauge symmetry and its spontaneous breaking--the origin of (right-handed) neutrino mass}
Given the new definition of $\overline C,P,T$ symmetries for a Majorana fermion, we are ready to discuss the origin of the neutrino mass, assuming that the neutrino is a Majorana fermion.
We can construct a mass term preserving time reversal symmetry, parity symmetry and spin rotational symmetry:
\begin{eqnarray}
H_m=\frac{m}{2}\left[i \gamma_\up(\v x) \gamma_{\up}^\prime(\v x)-i \gamma_\down(\v x) \gamma_{\down}^\prime(\v x)\right]
\end{eqnarray}
However, such a mass term breaks the charge conjugation symmetry since $\overline C H_m \overline C^{-1}=-H_m$.

If we elevate the charge conjugation symmetry to a $\mathbb{Z}_2$ gauge symmetry, the origin of the Majorana mass term can be explained as the spontaneous gauge symmetry breaking through the Anderson-Higgs mechanism\cite{higgs}.
The fundamental $\mathbb{Z}_2$ gauge field is potentially be detectable via cosmic string($\mathbb{Z}_2$ flux line) in the early universe. Finally, to be compatible with the SM, the neutrino mass discussed here should be the mass of the right-handed sterile neutrino, since a Majorana mass term for the left-handed light neutrino is not allowed in the original SM(no extension of the electroweak Higgs sector) and can only be induced through the seesaw mechanism\cite{seesaw1,seesaw2,seesaw3,seesaw4}.

To implement the above idea in quantum field theory, we can introduce a new real scalar field $\phi(x)=\phi(t,\v x)$ which carries $\mathbb{Z}_2$ gauge charge one(thus it transforms as $\overline C \phi(x) \overline C^{-1}=-\phi(x)$) and couple it to the Majorana field. The Anderson-Higgs mechanism\cite{higgs} can be realized by condensing the real scalar field $\phi(x)$. We assume that such a fundamental scalar field does not carry other gauge charge and is invariant under the $P$ and $T$ symmetry.
The following Lagrangian preserves all the $\overline C,P,T$ symmetries:
\begin{eqnarray}
\mathcal{L}&=&\mathcal{L}_0+\mathcal{L}_m+\mathcal{L}_\phi+\mathcal{L}_{\mathbb{Z}_2}\nonumber\\&=&\frac{1}{4}\overline\psi_c(x)i\gamma_\mu D_\mu \psi_c(x)+\frac{ig}{4}\phi(x)\overline\psi_c(x)\gamma_5\psi_c(x)\nonumber\\&+&|D_\mu \phi|^2-V(\phi)+\mathcal{L}_{\mathbb{Z}_2}
\end{eqnarray}

If we assume that the real scalar field condenses at $\langle\phi(x)\rangle=\phi_0$, a mass term $im\overline\psi_c(x)\gamma_5\psi_c(x)$ arises with $m=g\phi_0/4$. Here $D_\mu$ represents the covariant derivative and $\mathcal{L}_{\mathbb{Z}_2}$ represents the action of $\mathbb{Z}_2$ gauge field. (We need to regulate the field theory in a discrete space-time or use topological BF theory to write down its explicit form.)

\section{Origin of three generations of neutrinos}
\label{sec: three generation}
\subsection{General discussion and some physical pictures}
The existence of three generations of neutrinos is one of the biggest mysteries in our universe. In this section, we will show that such a puzzle can be naturally resolved by assuming that a Majorana fermion is made up of four topological Majorana zero modes. The key observation is that there are \emph{three} inequivalent ways to define a pair of Majorana spinons that describe a pair of complex fermions(with opposite spin polarizations) out of four topological Majorana zero modes. More precisely, the pair of Majorana spinons can be made up not only by $(\gamma_\up,\gamma_\down),(\gamma_\up^\prime,\gamma_\down^\prime)$, but also by
$(\gamma_\up^\prime,\gamma_\down),(\gamma_\up,\gamma_\down^\prime)$ or $(\gamma_\up,\gamma_\up^\prime),(\gamma_\down,\gamma_\down^\prime)$.

Let us define:
\begin{eqnarray}
d_L=\frac{1}{2}(\gamma_\up-i\gamma_\down^\prime); \quad d_R=\frac{1}{2}(\gamma_\up^\prime-i\gamma_\down),
\end{eqnarray}
Under the $\overline C,P,T$ symmetries, they transform as:
\begin{eqnarray}
\overline C d_L \overline C^{-1}&=&-i d_L; \quad \overline C d_R \overline C^{-1}=i d_R \nonumber\\
P d_L P^{-1}&=& -d_R; \quad P d_R P^{-1}=d_L \nonumber\\
T d_L T^{-1}&=&i d_R^\dagger; \quad T d_R T^{-1}=i d_L^\dagger,\label{halfZ2}
\end{eqnarray}
Similarly, we can define:
\begin{eqnarray}
f_L=\frac{1}{2}(\gamma_\up+i\gamma_\up^\prime)=c_\up; \quad f_R=\frac{1}{2}(\gamma_\down+i\gamma_\down^\prime)=c_\down^\dagger
\end{eqnarray}
Under the $\overline C,P,T$ symmetries, they transform as:
\begin{eqnarray}
\overline C f_L \overline C^{-1}&=&i f_R; \quad \overline C f_R \overline C^{-1}=i f_L \nonumber\\
P f_L P^{-1}&=&i f_L; \quad P f_R P^{-1}=-i f_R \nonumber\\
T f_L T^{-1}&=&-f_R^\dagger; \quad T f_R T^{-1}=f_L^\dagger,
\end{eqnarray}
We see that $d_{L(R)}$ and $f_{L(R)}$ fermions transform differently under the $\overline C,P,T$ symmetries. Especially, the local Fock space of $d_{L(R)}$ carries the $(TP)^4=-1$ projective representation of $TP$ symmetry while the local Fock space of $f_{L(R)}$ carries the $(T\overline C)^4=-1$ projective representation of $T \overline C$ symmetry. We have:
\begin{eqnarray}
(TP) d_L (TP)^{-1}&=&-i d_L^\dagger; \quad (TP) d_R (TP)^{-1}=i d_R^\dagger \nonumber\\
(T\overline C) f_L (T\overline C)^{-1}&=&-i f_L^\dagger; \quad (T\overline C) f_R (T\overline C)^{-1}=if_R^\dagger,
\end{eqnarray}
Apparently the above $TP$ and $T \overline C$ transformations for $d_{L(R)}$ and $f_{L(R)}$ fermions have the same form as Eq.(\ref{TLR}), therefore they carry the same representation theory as Eq.(\ref{RepT}).

From a condensed matter theory point of view, the above argument can be understood as there are three different types of point-like topological defects in a TSC protected by $\overline C, P$ and $T$ symmetries, characterized by the $T^4=-1$, $(TP)^4=-1$, and $(T\overline C)^4=-1$ projective symmetries that the corresponding topological Majorana zero modes carry. Since point-like defects can only be created/anihilated in pairs, there is a natural notion of neutrino and anti-neutrino pair. In the following, we again construct some explicit $1$D TSC models to further explain this idea.

\begin{figure}[t]
\begin{center}
\includegraphics[width=7cm]{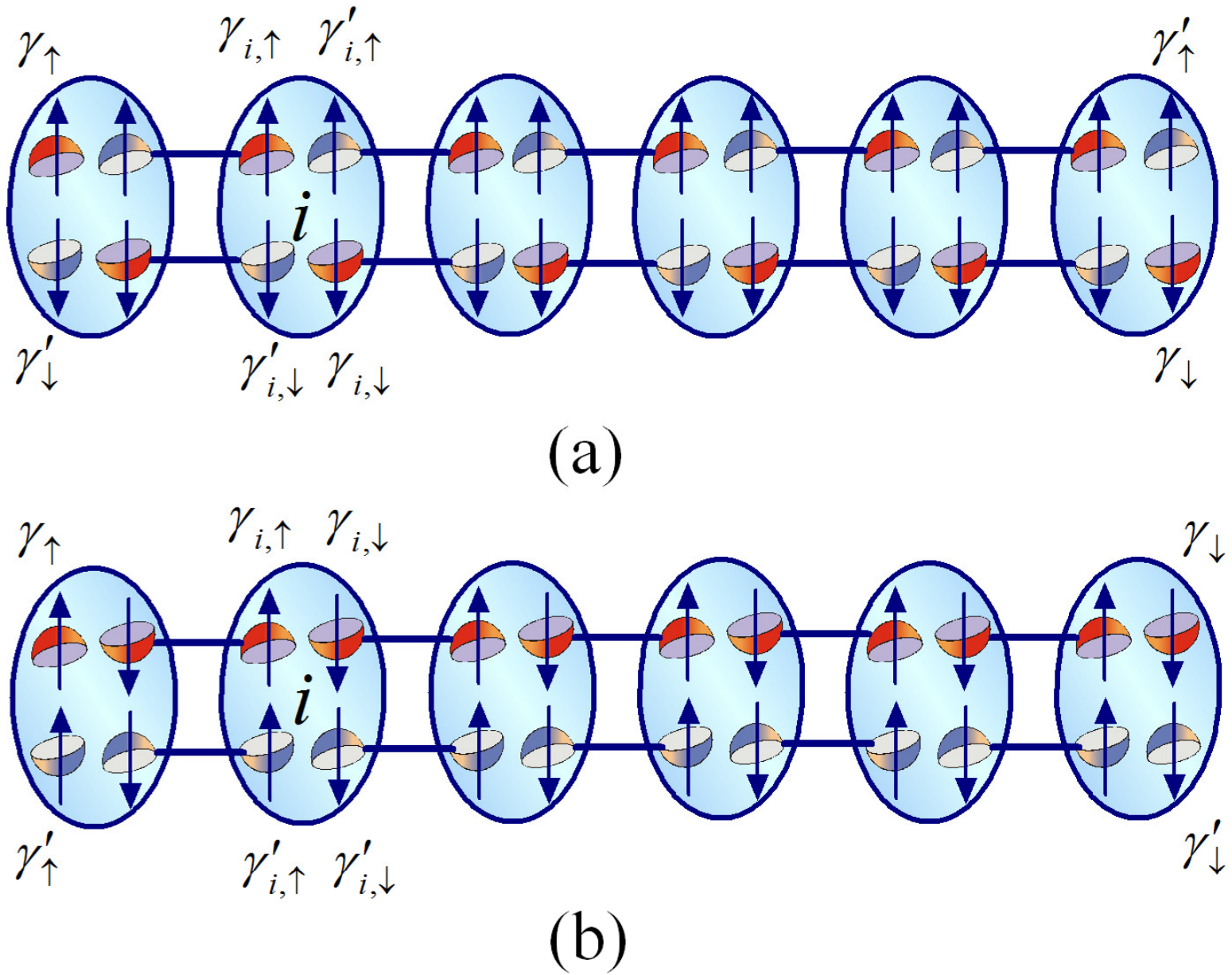}
\caption{(color online)The other two $1$D TSC models protected by $TP$ and $T\overline C$ symmetries, the topological Majorana modes on their ends carry the $(TP)^4=-1$ and $(T \overline C)^4=-1$ projective representations.
}\label{chain1}
\end{center}
\end{figure}
Similar to the time reversal protected Majorana chain that has been discussed at the very beginning of this paper, we can also construct $TP$(Here again we only consider the internal action of $P$ symmetry, since the symmetry protected nature of topological Majorana zero modes only relies on the internal action and has nothing to do with the coordinate action.) and $T\overline C$ protected Majorana chains explicitly. Let us consider the following Hamiltonian:
\begin{eqnarray}
H_d=\sum_{i=1}^N \left(i\gamma_{i,\up}^\prime\gamma_{i+1,\up}+i\gamma_{i,\down}\gamma_{i+1,\down}^\prime\right),\label{Hd}
\end{eqnarray}
and
\begin{eqnarray}
H_f=\sum_{i=1}^N \left(i\gamma_{i,\down}\gamma_{i+1,\up}+i\gamma_{i,\down}^\prime\gamma_{i+1,\up}^\prime\right),\label{Hf}
\end{eqnarray}
It is clear that $H_d$ is invariant under the $TP$ symmetry and $H_f$ is invariant under the $T\overline C$ symmetry. In Fig. \ref{chain1}, we see that for $H_d$, the pair of topological Majorana modes on both ends form a $(TP)^4=-1$ representation, while for $H_f$, the pair of topological Majorana modes on both ends form a $(T\overline C)^4=-1$ representation. All our discussions for the $1$D model can be generalized into $3$D as well, where the topological Majorana modes will be localized on the hedgehog/anti-hedgehog, and similar hedgehog/anti-hedgehog lattice model Eq. (\ref{3Dlattice}) with proliferated topological Majorana modes can be constructed in the same way by replacing $c_{L(R)}$ fermion with $d_{L(R)}$ and $f_{L(R)}$ fermions.

\subsection{Possible internal structure of Majorana fermion: a semiclassical picture}
\begin{figure}[t]
\begin{center}
\includegraphics[width=6.5cm]{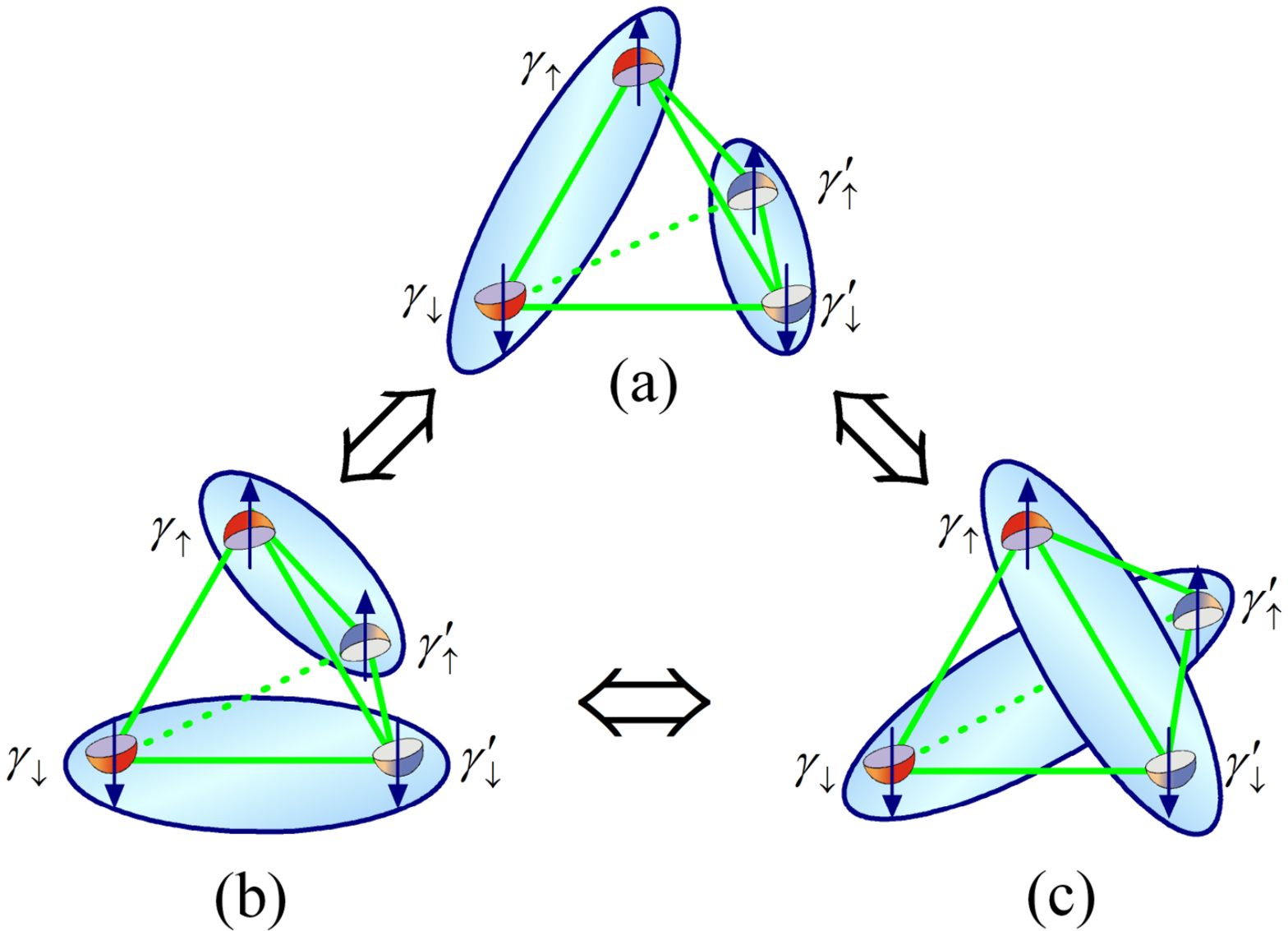}
\caption{(color online)A conjectured internal structure of a Majorana fermion at cutoff scale, which consists of four topological Majorana zero modes located on the vertices of a tetrahedra. A topological Majorana zero mode must be attached to the end of a (topological) open string.
The internal structure of neutrino suggests that the three generations of neutrinos/anti-neutrinos can be explained as three different ways of forming a pair of complex fermions out of four topological Majorana modes, characterized by the $T^4=-1$, $(TP)^4=-1$ and $(T \overline C)^4=-1$ symmetries that they carry.
}\label{internal}
\end{center}
\end{figure}

Although the lattice model of topological defects is very promising and insightful for us to understand the origin of three generations of neutrinos, it has been believed that a fundamental theory does not necessarily emerge from any pre-assumed lattice model. Here we would like to provide an alternative understanding for the origin of three generations of neutrinos by proposing a possible internal structure of a Majorana fermion. As seen in Fig. \ref{internal}, we conjecture that a Majorana fermion is actually made up of four topological Majorana zero modes located on the four vertices of a tetrahedra at cutoff scale. However, since a topological Majorana zero mode carries Non-abelian statistics and could not be a point-like particle in 3D, it must be attached to the end of a fundamental open string.
In such a physical picture, the origin of three generations of neutrinos can be explained by three different ways of forming a pair of complex fermions(with opposite spin polarizations) out of four topological Majorana modes, namely, $c_{L(R)}^\dagger $, $d_{L(R)}^\dagger $ and $f_{L(R)}^\dagger $, identified by the $T^4=-1$, $(TP)^4=-1$ and $(T \overline C)^4=-1$ symmetries that each complex fermion carries.

Indeed, both the internal structure and topological defect picture share the same spirit: the Hilbert space for each pair of topological Majorana modes must be spatially separated at cutoff scale to make the projective representations $T^4=-1$, $(TP)^4=-1$ and $(T \overline C)^4=-1$ meaningful.
Thus, the three generations of neutrinos can be uniquely identified by the fractionalized $\overline C, P, T$ symmetries that they carry at cutoff energy scale. We would like to stress that since a Dirac fermion can always be decomposed into a pair of Majorana fermions, the topological Majorana zero mode scenario will be applicable for the Dirac fermion as well. As a result, the origin of three generations of quarks and charged leptons can be understood in the same way.
Unfortunately, the above single particle picture can not be generalized into quantum field theory, since a rigorous way to incorporate the internal structure of a fundamental quantum field is absent so far.

\subsection{A potential topological quantum field theory description and flavor gauge symmetry at cut-off energy scale}
Before closing this section, we would like to mention a potential mathematical framework toward constructing a topological quantum field theory that can incorporate the internal structure of a Majorana fermion consisting of four topological Majorana modes. Comparing to the lattice regulation scheme proposed in Fig. \ref{3Dchain}, the TQFT regulation has its unique advantage of preserving Lorentz invariance, even at cutoff scale. In $2$D, it is well known that the (Ising) UMTC theory serves as an algebraic description of the underlying TQFT for topological Majorana modes. In such a framework, a single topological Majorana mode $\gamma$ actually represents a half degree of freedom with Non-Abelian statistics -- the Ising anyon $\sigma$, and the fusion rule of a $\sigma$ particle satisfies $\sigma\times\sigma=1+\psi$, where $\psi$ is a complex fermion. In the presence of $P$ and $T$ symmetries, there will be four kinds of topological Majorana modes $\gamma_\up$, $\gamma_\down$, $\gamma_\up^\prime$, $\gamma_\down^\prime$ representing four different Ising anyons $\sigma_\up$, $\sigma_\down$, $\sigma_\up^\prime$, $\sigma_\down^\prime$. They can fuse into complex fermions in three different ways.
\begin{eqnarray}
&&\sigma_\up\times \sigma_\down \times \sigma_\up^\prime \times \sigma_\down^\prime =
(1+\psi_c)(1+\psi_c^\prime)\\\nonumber &=&(1+\psi_d)(1+\psi_d^\prime)=(1+\psi_f)(1+\psi_f^\prime)
\end{eqnarray}
In principle, such kind of fusion rules can be generalized into $3$D as well. Nevertheless, as aforementioned, since point-like particle in $3$D can only carry boson or fermion statistics, anyon-like objects must be realized on the ends of a topological string.(The word "topological" means tensionless, e.g, a vortex line in a deconfined discrete gauge theory.) Therefore, we need to develop an algebraic theory for string-like extensive topological objects. Recently, it has been shown that strings in $3$D can carry nontrivial statistics\cite{3Dstring1,3Dstring2}, and a potential algebraic theory would be the mathematically unknown UMT2C theory, which will obviously be an attractive future direction.

Another advantage of the above TQFT framework is that it naturally unifies three generations of neutrinos at the topological string energy scale. In the semiclassical picture Fig. \ref{internal}, the three different pairs of topological Majorana modes should be thought as three different types of (topological) open strings and they can fuse into each other. Therefore, at the TQFT scale, three generations of neutrinos will be unified and they can be viewed as three different resonating valence bonds(RVB) formed by topological Majorana modes.

In other words, at cutoff energy scale, we can view the flavor symmetry as a gauge symmetry in the fundamental TQFT and three generations of neutrinos are indeed three different local gauge choices to label the same Hilbert space expanded by four topological Majorana modes.(We note that "local gauge symmetry" is actually not a symmetry but a relabeling scheme of the same Hilbert space.) At low energy, e.g., in the SM, the flavor gauge symmetry will be spontaneously broken and we can introduce three independent fields to describe the low energy physics. In terms of physical picture, one can view the SM as the string confinement phase and each tiny string in Fig. \ref{internal} becomes point-like particle carrying different fractionalized $\overline C, P, T$ symmetries.(We note that the string confinement phase naturally introduces an energy scale of string tension and it is no longer a TQFT.) More precisely, in a TQFT, e.g., a topological BF theory $S_{\text{top}}=\frac{1}{2\pi}\int B\wedge F$, has an enhanced $SL(4,R)$ spacetime symmetry and a four component Majorana fermion consisting of four topological Majorana zero modes is indeed a $SL(4,R)$ spinor. Thus, the string confinement phase will lead to a spontaneous symmetry breaking of $SL(4,R)$ spacetime symmetry down to Lorentz symmetry $SO(3,1)$ with $SL(2,C)$ spinor. In fact, there are three and only three different ways of such a spontaneous symmetry breaking since $SL(4,R)$ is isomorphic to $SO(3,3)$, and its breaking down to $SO(3,1)$ can be realized by choosing one of the time-like directions in $SO(3,3)$. 

In next Section, we will first develop a "poor man" quantum field theory approach by taking the continuum limit of the lattice model proposed in Fig. \ref{3Dchain} and use a semiclassical approximation to compute the neutrino mass mixing matrix and neutrino mass ratios among three generations. Then we will use the concept of flavor gauge symmetry to derive the same results. These predictions can be carefully examined by future experiments.

\section{Application: right-handed neutrino mass mixing matrix and $A_5$ flavor symmetry}
\label{sec: mass mixing}
\subsection{Seesaw mechanism}
It is well known that a Majorana mass term of the form $m\overline\psi(x)\psi(x)$ is prohibited for left-handed light neutrinos since it breaks the electroweak gauge symmetry, and that is why the original SM predicts zero neutrino mass. A nice way to fix this problem is to assume the existence of three generations of heavy sterile neutrinos, and masses of the left-handed light neutrinos can be induced by the type-I seesaw mechanism\cite{seesaw1,seesaw2,seesaw3,seesaw4}. The total mass matrix reads:
\begin{eqnarray}
 M_{total}=\left(
   \begin{array}{cc}
     0 & m_D \\
     m_D & M \\
   \end{array}
 \right),
\end{eqnarray}
where $m_D$ is the 3 by 3 Dirac mass matrix and $M$ is the 3 by 3 Majorana mass matrix of right-handed sterile neutrinos.(We note that the left-handed neutrinos have a zero mass.) If we assume that $m_D$ is at the electroweak symmetry breaking energy scale($250GeV$) and $M$ is at the grand unification theory(GUT) energy scale($10^{15}GeV$), a mass at the energy scale of $0.1eV$ can be induced for the left-handed light neutrinos.
There is no general principle to fix $m_D$ as well as the mass matrix of charged leptons in SM, unless we impose certain flavor symmetry within LO approximation(e.g., in the absence of CP violation correction).

We will first apply the idea of topological Majorana zero modes to derive the right-handed neutrino mass matrix, which essentially implies an $A_5$ flavor symmetry. Then we apply the $A_5$ flavor symmetry, together with the derived right-handed neutrino mass matrix to compute the mass mixing matrix as well as exact mass ratios for left-handed neutrinos within LO approximation.
The potential topological origin of the $A_5$ flavor symmetry will be discussed at the end of this section.

\subsection{A "poor man" quantum field theory description for three generations of right-handed neutrinos at cutoff energy scale}
In order to derive the right-handed neutrino mass matrix $M$ in extended SM, here we would like to develop a "poor man" quantum field theory description for three generations of neutrinos. As having been pointed out in section \ref{sec: three generation}, without an explicit cutoff, there is no way to distinguish the three generations of Majorana fermions made up of four topological Majorana zero modes.
However, if we have already introduced three independent Majorana fields in SM, there is no difficulty for us to develop a "poor man"  quantum field theory to describe their unusual $\overline C, P, T$ properties, e.g., by taking the continuum limit of the lattice model Fig. \ref{3Dchain}, and for the purpose of computing $M$ semiclassically, such a "poor man" approach would be sufficient.

The key idea of constructing the lattice model Fig. \ref{3Dchain} is to split a four dimensional Fock space expanded by $c_{\v x,\up}^\dagger |0\rangle,c_{\v x,\down}^\dagger|0\rangle, |0\rangle,c_{\v x,\up}^\dagger c_{\v x,\down}^\dagger|0\rangle$ at a single spacial point $\v x$ into a pair of two dimensional Hilbert spaces on each single site at cutoff scale(assuming there are two sublattices per unit cell). This is possible and natural since a spacial point $\v x$ represents a unit cell at low energy and long wavelength. As seen in Fig. \ref{3Dchain}, to describe the $c_{L(R)}$ fermion, we just need to put a pair of topological Majorana modes $\gamma_\up,\gamma_\down$ on sublattice-A while another pair $\gamma_\up,\gamma_\down$ on sublattice-B. It is clear that in such a construction, the Majorana spinon basis $\xi(\v x_A)$ and $\eta(\v x_B)$ carrying $T^4=-1$ time reversal symmetry correspond to the local degrees of freedom on sublattice-A and sublattice-B. Only at long wavelength and low energy when $\v x_A$ and $\v x_B$ are identified as the same spacial point $\v x$, the four component relativistic Majorana field $\psi_c(\v x)$ emerges.

Alternatively, we can also put topological Majorana zero modes $(\gamma_\up^\prime,\gamma_\down),(\gamma_\up,\gamma_\down^\prime)$ or $(\gamma_\up,\gamma_\up^\prime),(\gamma_\down,\gamma_\down^\prime)$ on sublattice-A and sublattice-B, corresponding to the $(TP)^4=-1$ and $(T\overline C)^4=-1$ projective representations on each sublattice.
To describe neutrino(anti-neutrino) made by $f_{L(R)}$ fermion at cutoff energy scale, we just need to define the Majorana fermion field $\psi_f(x)= \left(
\begin{array}{c}
   \t\xi(x)\\
       \t\eta(x) \\
\end{array}
\right) $ with a different Majorana spinon basis:
\begin{eqnarray}
\t\xi(x)= \left(
        \begin{array}{c}
          \t \gamma_\up(x)  \\
          \t \gamma_\up^\prime(x)  \\
        \end{array}
      \right);\quad
\t\eta(x)= \left(
        \begin{array}{c}
          \t \gamma_\down(x) \\
          \t \gamma_\down^\prime(x)  \\
        \end{array}
      \right),
\end{eqnarray}
The above Majorana fermion field satisfies the $\overline C,P,T$ symmetries:
\begin{eqnarray}
 \overline C \psi_f(x) \overline C^{-1}&=&-\gamma_5\psi_f(x); \quad P  \psi_f(x) P^{-1}= \gamma_0 \psi_f(\t x);\nonumber\\ T \psi_f(x) T^{-1}&=&-\gamma_0\gamma_5\psi_f(-\t x),
\end{eqnarray}
where the definitions of $\overline C,P,T$ operators are the same as Eq.(\ref{CPT}) after we replace all the $\gamma_\sigma(\v x), \gamma_\sigma^\prime(\v x)$ by $\t\gamma_\sigma(\v x), \t\gamma_\sigma^\prime(\v x)$.
It is clear that the $f_{L(R)}$ fermion transforms differently under $\overline C,P,T$ symmetries, and for the $f_{L(R)}$ fermion, its mass term takes the usual form:
\begin{eqnarray}
\mathcal{L}_m=\frac{ig}{4}\phi(x)\overline\psi_f(x)\psi_f(x),\quad \overline\psi_f(x)= \psi_f^\dagger (x) \gamma_0
\end{eqnarray}

Finally, for the neutrino(anti-neutrino) made by the $d_{L(R)}$ fermion at cutoff energy scale, we need to choose $\bar\gamma_0=R\gamma_0 R^{-1}=i\rho_x\otimes\sigma_y\equiv\gamma_5$ with:
\begin{eqnarray}
R=\frac{1}{\sqrt{2}}\left(
        \begin{array}{cc}
          1 & 1 \\
          -1 & 1 \\
        \end{array}
      \right)=\frac{1}{\sqrt{2}}(1+\gamma_0\gamma_5)
\end{eqnarray}
The corresponding $\gamma_{1,2,3}$ and  $\gamma_5$ transform as:
$\bar \gamma_{1,2,3}=R\gamma_{1,2,3}R^{-1}=\gamma_{1,2,3}$ and $\bar\gamma_5=R\gamma_5 R^{-1}=i\rho_z\otimes\sigma_y\equiv-\gamma_0$). Indeed, this representation was first proposed by Ettore Majorana.

The quantum field theory can be obtained by defining $\psi_d(x)= \left(
\begin{array}{c}
   \hat\xi(x)\\
       \hat\eta(x) \\
\end{array}
\right)$ with:
\begin{eqnarray}
\hat\xi(x)= \left(
        \begin{array}{c}
          \hat \gamma_\up(x)  \\
          \hat \gamma_\down^\prime(x)  \\
        \end{array}
      \right);\quad
\hat\eta(x)= \left(
        \begin{array}{c}
          \hat \gamma_\down(x)  \\
          -\hat \gamma_\up^\prime(x) \\
        \end{array}
      \right),
\end{eqnarray}
Under the $\overline C, P, T$ symmetries with above definition, $\psi_d(x)$ transforms as:
\begin{eqnarray}
 \overline C \psi_d(x)  {\overline C}^{-1}&=&-\bar\gamma_5\psi_d(x)\equiv\gamma_0\psi_d(x); \nonumber\\ P \psi_d(x) {P}^{-1}&=& \bar\gamma_0\psi_d(\t x)\equiv\gamma_5\psi_d(\t x);\nonumber\\
 T \psi_d(x) {T}^{-1}&=&-\bar\gamma_0\bar\gamma_5\psi_d(-\t x)\equiv-\gamma_0\gamma_5\psi_d(-\t x),
\end{eqnarray}
where the definitions of $\overline C,P,T$ operators are also the same as Eq.(\ref{CPT}) after we replace all the $\gamma_\sigma(\v x), \gamma_\sigma^\prime(\v x)$ by $\hat\gamma_\sigma(\v x), \hat\gamma_\sigma^\prime(\v x)$.
For the $d_{L(R)}$ fermion, the mass term also takes the usual form:
\begin{eqnarray}
\mathcal{L}_m=\frac{ig}{4}\phi(x)\overline\psi_d(x)\psi_d(x),\quad \overline\psi_d(x)= \psi_d^\dagger (x)\bar \gamma_0=\psi_d^\dagger (x) \gamma_5,\nonumber\\ \label{dfield}
\end{eqnarray}

The three generations of neutrino fields described by $c_{L(R)}$, $f_{L(R)}$ and $d_{L(R)}$ fermions can also be identified by their different $\overline C,P,T$ transformation laws in momentum space, see Appendix \ref{App:momentum} for details.

We would like to argue that the above "poor man" quantum field theory descriptions are very general and do not depend on any particular scheme of lattice regulation. Actually, due to the fermion doubling problem, the sublattice structure introduced in the lattice model Fig. \ref{3Dchain} can not be avoided. Therefore, we can in principle construct a dynamical lattice model and restore the Lorentz symmetry even at cutoff scale. As long as the sublattice structure is imposed, we will end up with the same continuum field theory descriptions for three generations of right-handed neutrinos. 

To this end, let us clarify the key difference between extended SM and our "poor man" quantum field theory descriptions for three generations of right-handed neutrinos. In SM. three generations of right-handed neutrinos are described as three copies of the same right-handed Weyl fermion fields, and there is not any quantum number that can distinguish them. However, in our "poor man" quantum field theory descriptions, we use three different pairs of Majorana spinon basis $\xi(\v x),\eta(\v x)$, $\t \xi(\v x),\t \eta(\v x)$ and $\hat \xi(\v x),\hat \eta(\v x)$ to describe three generations of righ-handed neutrinos. If we are not allowed to redefine the four component Majorana fields $\psi_c$, $\psi_f$ and $\psi_d$ by mixing the pair of Majorana spinon basis $\xi(\v x),\eta(\v x)$, $\t \xi(\v x),\t \eta(\v x)$ and $\hat \xi(\v x),\hat \eta(\v x)$, they can be distinguished by different $\overline C, P, T$ properties. Apparently, this is a reasonable assumption for right-handed neutrino before they get a mass, since $M$ is very close to the cut-off scale. 
On the other hand, at low energy, e.g., in the extended SM where right-handed neutrinos has already got a very big mass, we should allow field redefinition for the full 4-component Majorana fields $\psi_c(x)$, $\psi_d(x)$ and $\psi_f(x)$. Thus, we can make them have the same $\overline C, P, T$ properties, which is the standard convention in extended SM.

In section \ref{sec: field theory}, we propose that the origin of the right-handed neutrino mass can be understood as the spontaneous breaking of the $\mathbb{Z}_2$ charge conjugation gauge symmetry. In the following we will apply the same idea to derive the entire right-handed neutrino mass matrix $M$.

\subsection{Right-handed neutrino mass matrix}
First, according to the $\mathbb{Z}_2$ gauge (minimal coupling) principle, we can write down the most general $\overline C,P,T$ invariant mass term for three generations of right-handed neutrinos. We have:
\begin{widetext}
\begin{eqnarray}
\mathcal{L}_m &=&\frac{ig}{4}\phi(x)\left[\overline {\psi}_f(x)\psi_f(x)+\overline{\psi}_d(x)\psi_d(x)+\overline{\psi}_c(x)\gamma_5\psi_c(x)\right]\nonumber\\
&+&\frac{ig^\prime}{4}\phi(x)\left[\overline {\psi}_d(x)(1+\gamma_0\gamma_5)(1+\gamma_5)\psi_c(x)+\overline {\psi}_c(x)(1+\gamma_5)(1-\gamma_0\gamma_5)\psi_d(x)\right]
\nonumber\\&+&\frac{ig^\prime}{4}\phi(x)\left[\overline{\psi}_f(x)(1+\gamma_5)\psi_c(x)
+\overline\psi_c(x)(1+\gamma_5)\psi_f(x)\right]\nonumber\\&+&\frac{ig^\prime}{4}\phi(x)\left[\overline{\psi}_f(x)(1-\gamma_0\gamma_5)\psi_d(x)
+\overline\psi_d(x)(1+\gamma_0\gamma_5)\psi_f(x)\right]\label{mass}
\end{eqnarray}
\end{widetext}
Here we use the same coupling $g$ for all the diagonal mass terms and $g^\prime$ for all the off-diagonal mass terms. Again, this is because the three generations of right-handed neutrinos are the three resonating states out of the \emph{same} four topological Majorana zero modes at cutoff scale. The above argument can also be incorporated
into quantum field theory language(in the absence of cutoff physics) by imposing a flavor gauge symmetry to constrain the coupling constant, see next subsection for details.  We note that for $\psi_c(x)$ and $\psi_f(x)$, the boost generators are defined by $S_{0i}=\frac{1}{4}[\gamma_0,\gamma_i]$ while for $\psi_d(x)$, the boost generator is defined by $\bar S_{0i}=\frac{1}{4}[\bar\gamma_0,\gamma_i]=\frac{1}{4}[\gamma_5,\gamma_i]$. Such an interesting twist makes the above mass term invariant under the Lorentz transformation, despite the existence of $(1\pm\gamma_0\gamma_5)$ term which does not seem to be invariant under the Lorentz boost.

In the extended SM(where $\phi(x)\sim \phi_0$ is condensed), three generations of right-handed neutrinos are described by three copies of the same Majorana field. Let us redefine $\psi_f(x)$ by:
\begin{eqnarray}
\psi_f^\prime(x)\equiv R\psi_f(x)\equiv\frac{1}{\sqrt{2}}(1+\gamma_0\gamma_5)\psi_f(x),
\end{eqnarray}
The corresponding $\gamma_0$ and $\gamma_{1,2,3}$ will change into $ \bar\gamma_0$ and $ \bar\gamma_{1,2,3}$.
Similarly, we can also redefine $\psi_{c}$ by:
\begin{eqnarray}
\psi_{c}^\prime(x)\equiv R \frac{1+\gamma_5}{\sqrt{2}}\psi_{c}(x)\equiv \frac{1}{2}(1+\gamma_0\gamma_5)({1+\gamma_5})\psi_{c}(x),\nonumber\\
\end{eqnarray}
It is easy to see that $\psi_f^\prime(x)$, $\psi_d(x)$  and $\psi_{c}^\prime(x)$ transform in the same way under the $\overline C,P,T$ symmetries, therefore, they can be interpreted as the three generations of right-handed sterile neutrinos in the extended SM. In terms of $\psi_f^\prime(x)$, $\psi_d(x)$ and $\psi_c^\prime(x)$, the $\overline C,P,T$ invariant mass term takes the following form:
\begin{eqnarray}
\mathcal{L}_m &=&\frac{ig}{4}\phi_0\left[\overline {\psi}_f^\prime(x)\psi_f^\prime(x)+\overline{\psi}_d (x)\psi_d(x)+\overline{\psi^\prime}_c(x)\psi^\prime_c(x)\right]\nonumber\\&+&\frac{2ig^\prime}{4}\phi_0\left[\overline {\psi}_d(x)\psi_c^\prime(x)+\overline {\psi^\prime}_c(x)\psi_d(x)\right]
\nonumber\\&+&\frac{\sqrt{2}i g^\prime}{4}\phi_0\left[\overline{\psi^\prime}_f(x)\psi_c^\prime(x)
+\overline{\psi^\prime}_c(x)\psi^\prime_f(x)\right]\nonumber\\&+&\frac{\sqrt{2}ig^\prime}{4}\phi_0\left[\overline{\psi^\prime}_f(x){\psi}_d(x)
+\overline{\psi}_d(x)\psi^\prime_f(x)\right]
\end{eqnarray}
with $\overline\psi^\prime_{f}(x)=(\psi^\prime_{f})^\dagger(x)\bar\gamma_0$ and $\overline\psi^\prime_{c}(x)=(\psi^\prime_{c})^\dagger(x)\bar\gamma_0$. (We note that the corresponding boost generators should be $\bar S_{0i}=\frac{1}{4}[\bar\gamma_0,\gamma_i]$.)

We see that the mass mixing pattern has already been fixed, regardless of the relative strength of $g$ and $g^\prime$. The mass matrix can be diagonalized by(the basis is ordered as $\psi^\prime_f,\psi_d,\psi_c^\prime$ and $\frac{\phi_0}{4}$ is set to be $1$):
\begin{eqnarray}
M&=&\left(
    \begin{array}{ccc}
      g &  \sqrt{2}g^\prime &  \sqrt{2}g^\prime \\
      \sqrt{2}g^\prime & g &  2g^\prime \\
       \sqrt{2}g^\prime &  2g^\prime & g \\
    \end{array}
  \right)\\&=&V^\dagger \left(
                      \begin{array}{ccc}
                        (1-\sqrt{5})g^\prime+g & 0 & 0 \\
                        0 & (1+\sqrt{5})g^\prime+g & 0 \\
                        0 & 0 & -2g^\prime+g \\
                      \end{array}
                    \right)V,\nonumber \label{diag}
\end{eqnarray}
with
\begin{eqnarray}
V^\dagger&=&\left(
                     \begin{array}{ccc}
                                             \sqrt{\frac{5+\sqrt{5}}{10}} &    \sqrt{\frac{5-\sqrt{5}}{10}} &0 \\
                     -\sqrt{\frac{5-\sqrt{5}}{20}} & \sqrt{\frac{5+\sqrt{5}}{20}} & -\frac{1}{\sqrt{2}} \\
                       -\sqrt{\frac{5-\sqrt{5}}{20}}& \sqrt{\frac{5+\sqrt{5}}{20}} & \frac{1}{\sqrt{2}} \\
                      \end{array}
                    \right)\nonumber\\&\simeq&
                    \left(
                     \begin{array}{ccc}
                        0.85 & 0.53 & 0\\
                        -0.37 & 0.6 &  -0.71 \\
                        -0.37 & 0.6 & 0.71 \\
                      \end{array}
                    \right)\label{mixing}
\end{eqnarray}
In terms of mixing angle, we have:
\begin{eqnarray}
\theta_{23}=-45^\circ;\quad \theta_{13}=0;\quad \theta_{12}=31.7^\circ=\arctan(\frac{\sqrt{5}-1}{2}),\nonumber\\ \label{angle}
\end{eqnarray}
We note that the physical masses of the mass egienstates are the absolute value of Eq.(\ref{diag}), with $M_1=|(1-\sqrt{5})g^\prime+g|$, $M_2=|(1+\sqrt{5})g^\prime+g|$ and $M_3=|-2g^\prime+g|$, and the $\pm$ sign in front of $\theta_{23}$ is just a gauge choice of the basis.
Finally, due to the same reason that the three generations of right-handed neutrinos are the three resonating states out of the \emph{same} four topological Majorana zero modes at cutoff scale, we further argue that the diagonal Yukawa coupling must have the same strength as the off-diagonal coupling with $|g|=|g^\prime|$.

\subsection{Computing the right-handed neutrino mass matrix by using the flavor gauge symmetry at cut-off energy scale}
\label{App:symmetry}
In this subsection, we provide a flavor gauge symmetry argument for the choice of Yukawa couplings in Eq.(\ref{mass}). The proposed internal structure of a Majorana fermion suggests that the flavor symmetry should be a gauge symmetry rather than a global symmetry at cutoff scale. Since a gauge symmetry is nothing but a relabeling of the same Hilbert space, we can replace $\t\gamma_\sigma,\hat\gamma_\sigma$ with $\gamma_\sigma$ and
$\t\gamma_\sigma^\prime,\hat\gamma_\sigma^\prime$ with $\gamma_\sigma^\prime$ at cutoff scale. This will provide us an alternative way to compute the mass matrix of right-handed sterile neutrinos. Such an approach could work because the mass mixing phenomenon for right-handed sterile neutrinos occurs at GUT scale, which is much higher than the extended SM energy scale that breaks flavor gauge symmetry.

Let us start with the diagonal term and assume there are three independent couplings $g_{f}$, $g_d$ and $g_c$.
\begin{widetext}
\begin{eqnarray}
\mathcal{L}_{m-diag} =\frac{i}{4}\phi(x)\left[g_f\overline {\psi}_f(x)\psi_f(x)+g_d\overline{\psi}_d(x)\psi_d(x)+g_c\overline{\psi}_c(x)\gamma_5\psi_c(x)\right]\label{massdiag}
\end{eqnarray}

At cutoff scale, all the mass terms should be regarded as interactions between the scalar particle $\phi$ and the four topological Majorana modes $\gamma_\up,\gamma_\down,\gamma_\up^\prime$ and $\gamma_\down^\prime$. For example, all the three terms in Eq.(\ref{massdiag}) can be expressed as:
\begin{eqnarray}
\frac{ig_f}{4}\phi(x)\overline {\psi}_f(x)\psi_f(x)&=&\frac{ig_f}{2}\phi(x)\left[\gamma_\up(x)\gamma_\up^\prime(x)-\gamma_\down(x)\gamma_\down^\prime(x)\right];\nonumber\\
\frac{ig_d}{4}\phi(x)\overline{\psi}_d(x)\psi_d(x)&=&
\frac{ig_d}{2}\phi(x)\left[\gamma_\up(x)\gamma_\up^\prime(x)-\gamma_\down(x)\gamma_\down^\prime(x)\right];\nonumber\\
\frac{ig_c}{4}\phi(x)\overline{\psi}_c(x)\gamma_5\psi_c(x)&=&
\frac{ig_c}{2}\phi(x)\left[\gamma_\up(x)\gamma_\up^\prime(x)-\gamma_\down(x)\gamma_\down^\prime(x)\right],
\end{eqnarray}
The above expression implies that the three mass terms are indeed the same term at cutoff. Physically,
we can attribute the existence of three generations of neutrinos to the three different (local) ways of making a pair of complex fermions(with opposite spin polarization) out of four topological Majorana zero modes.

The same argument also applies to the off-diagonal mass term:
\begin{eqnarray}
\mathcal{L}_{m-offdiag}
&=&\frac{ig_{cd}}{4}\phi(x)\left[\overline {\psi}_d(x)(1+\gamma_0\gamma_5)(1+\gamma_5)\psi_c(x)+\overline {\psi}_c(x)(1+\gamma_5)(1-\gamma_0\gamma_5)\psi_d(x)\right]
\nonumber\\&+&\frac{ig_{cf}}{4}\phi(x)\left[\overline{\psi}_f(x)(1+\gamma_5)\psi_c(x)
+\overline\psi_c(x)(1+\gamma_5)\psi_f(x)\right]\nonumber\\&+&\frac{ig_{df}}{4}\phi(x)\left[\overline{\psi}_f(x)(1-\gamma_0\gamma_5)\psi_d(x)
+\overline\psi_d(x)(1+\gamma_0\gamma_5)\psi_f(x)\right]\label{massod},
\end{eqnarray}
which can be expressed as:
\begin{eqnarray}
 &&\frac{ig_{cd}}{4}\phi(x)\left[\overline {\psi}_d(x)(1+\gamma_0\gamma_5)(1+\gamma_5)\psi_c(x)+\overline {\psi}_c(x)(1+\gamma_5)(1-\gamma_0\gamma_5)\psi_d(x)\right]\nonumber\\&=&
 \frac{ig_{cd}}{2}\phi(x)\left[\gamma_\up(x)\gamma_\up^\prime(x)-\gamma_\down(x)\gamma_\down^\prime(x)\right];
 \end{eqnarray}
 \begin{eqnarray}
 &&\frac{ig_{cf}}{4}\phi(x)\left[\overline{\psi}_f(x)(1+\gamma_5)\psi_c(x)
+\overline\psi_c(x)(1+\gamma_5)\psi_f(x)\right]\nonumber\\&=&
 \frac{ig_{cf}}{2}\phi(x)\left[\gamma_\up(x)\gamma_\up^\prime(x)-\gamma_\down(x)\gamma_\down^\prime(x)\right];
  \end{eqnarray}
 \begin{eqnarray}
 &&\frac{ig_{df}}{4}\phi(x)\left[\overline{\psi}_f(x)(1-\gamma_0\gamma_5)\psi_d(x)
+\overline\psi_d(x)(1+\gamma_0\gamma_5)\psi_f(x)\right]\nonumber\\&=&
 \frac{ig_{df}}{2}\phi(x)\left[\gamma_\up(x)\gamma_\up^\prime(x)-\gamma_\down(x)\gamma_\down^\prime(x)\right],
\end{eqnarray}
\end{widetext}
Thus we can derive $g_{cd}=g_{cf}=g_{df}=g^\prime$. Finally, by comparing the diagonal and off-diagonal mass terms, we can further derive $|g|=|g^\prime|$. Here the relative sign of $g$ and $g^\prime$ can not be fixed because a $Z_2$ gauge flux is possible for a loop enclosed by three generation hopping phases.


\subsection{$A_5$ flavor symmetry and prediction of left-handed neutrino mass}
\label{sec: mass symmetry}
 In the above, we use a semiclassical approach based on our topological scenario to compute the right-handed neutrino mass matrix(under certain basis choice). However, in order to make measurable predictions for the mass matrix of the left-handed neutrino, we must impose a proper flavor symmetry to determine the charged-lepton and the neutrino Yukawa couplings as well(We note that these Dirac mass terms at electroweak energy scale have a very different physical origin ).
 
 Since the mixing angle of right-handed neutrinos is consistent with the GR pattern\cite{masssymmetry1,masssymmetry2}, it is straightforward to check that $M$ is invariant under a $\mathbb{Z}_2\otimes \mathbb{Z}_2$ Klein symmetry with generators $U$ and $S$ defined by:
\begin{eqnarray}
U&=&\left(
    \begin{array}{ccc}
      1 & 0 & 0 \\
      0 & 0 & 1 \\
      0 & 1 & 0 \\
    \end{array}
  \right);
S=\frac{1}{\sqrt{5}}\left(
    \begin{array}{ccc}
      1 & -\sqrt{2} & -\sqrt{2} \\
      -\sqrt{2} & -\frac{(\sqrt{5}+1)}{2} & \frac{(\sqrt{5}-1)}{2} \\
      -\sqrt{2} & \frac{(\sqrt{5}-1)}{2}  &-\frac{(\sqrt{5}+1)}{2}\\
    \end{array}
  \right),\nonumber\\
\end{eqnarray}

In Ref.\cite{masssymmetry1}, it has been shown that the above $\mathbb{Z}_2\otimes \mathbb{Z}_2$ symmetry arises from an underlying $A_5$ flavor symmetry and that $S$ is one of the generators of $A_5$ group. Therefore, within LO approximation, it is natural to assume such an $A_5$ flavor symmetry,  which can further enforce the mass matrix of charged lepton to be diagonal and $m_D$ proportional to $U$\cite{masssymmetry1}. In fact, our topological scenario also gives rise to a natural origin of such $A_5$ flavor symmetry. This is because the spontaneous symmetry breaking of $SL(4,R)$ spacetime symmetry down to $SO(3,1))$ spacetime symmetry allows additional unbroken discrete symmetry, e.g. $A_5$. Thus, we can have a slightly modified symmetry breaking scheme from $SL(4,R)$ down to $SO(3,1)\otimes A_5$, which naturally explains the origin of $A_5$ flavor symmetry. In terms of physical picture, the above math statement can be understood as following: Although there are three and only three different ways to break $SL(4,R)$ spacetime symmetry down to $SO(3,1)$ spacetime symmetry(this explains the origin of three generations), an additional discrete $A_5$ remainant symmetry is still allowed in such a symmetry breaking pattern and it plays a role as flavor symmetry.

According to the seesaw mechanism, the mass mixing matrix for left-handed light neutrino takes the same form as Eq.(\ref{mixing})(in the limit $m_D\ll M$); however, the mass hierarchy is reversed.
Thus, the solution with $g=-g^\prime$ implies $M_1=M_2=\sqrt{5}g$ and $M_3=3g$, which leads to $m_1/m_3=m_2/m_3=3/\sqrt{5}$(here $m_1,m_2$ and $m_3$ are eigen masses of the left-handed light neutrinos) and can match the current experimental observations.(We assume that the small mass splitting $\Delta m_{12}$ is negligible within LO approximation.) On the contrary, the solution $g=g^\prime$ leads to $M_1=(\sqrt{5}-2)g<M_3=g<M_2=(\sqrt{5}+2)g$ and contradicts to the current experimental results with either $m_1\simeq m_2<m_3$ or $m_1\simeq m_2>m_3$. Therefore, here we choose $g=-g^\prime$.
 Based on the current experimental data $\Delta m^2_{23}\simeq 2.5\times10^{-3}eV^2$, we obtain $m_1=m_2\simeq0.075eV$ and $m_3\simeq0.054eV$.
In addition, we predict that $m_{0\nu\beta\beta}\equiv |\sum_i U_{ei}^2 m_i|=m_1/\sqrt{5}\simeq0.0335eV$.

Finally, we stress that in the usual $A_5$ flavor symmetry scenario, there is no constraint on mass ratios. Actually, the choice of $g=-g^\prime$ in our case leads to an additional $\mathbb{Z}_2$ symmetry generator $R$:
\begin{eqnarray}
 R=\frac{1}{\sqrt{2}}\left(
    \begin{array}{ccc}
      0 & i & i \\
      -i& \frac{1}{\sqrt{2}}  & -\frac{1}{\sqrt{2}} \\
      -i & -\frac{1}{\sqrt{2}} & \frac{1}{\sqrt{2}} \\
    \end{array}
  \right),
\end{eqnarray}

Together with $U$ and $S$, we find:
\begin{eqnarray}
U^TMU=M; \quad S^TMS=M,
\end{eqnarray}
and
\begin{eqnarray}
U^2&=&1;\quad S^2=1;\quad R^2=1,\nonumber \\ US&=&SU;\quad UR=RU;\quad SR=-URS,
\end{eqnarray}
which form a $D_4$ group. Since $D_4$ is not a subgroup of $A_5$, our scenario suggests an enlarged flavor symmetry which contains $A_5$ as a subgroup. For example, $S_5$ is a possible candidate since $R$ can be viewed as an additional reflection symmetry. Within LO approximation, the physical origin of the underlying flavor symmetry is a very deep and interesting problem. In fact, the $S_5$ symmetry group has a deep relationship with $3+1$D TQFT, which is the symmetry group of a $4$-simplex. In $2+1$D, it is well known that a large class of TQFT constructed from Turaev-Viro State-Sum invariants admits the full tetrahedra symmetry $S_4$ for the $6j$ $G$ symbol. In $3+1$D, a similar construction naturally admits an $S_5$ or $A_5$ symmetry for the corresponding $15j$ symbol.

\subsection{The effect of $CP$ violation}
Before conclusion, we discuss the effect of $CP$ violation for the neutrino mass mixing matrix. Recently, the DaYa-Bay's experiment has reported a nonzero $\theta_{13}\simeq 8.8^\circ$\cite{neutrinoexp9}. From our point of view, the experimentally observed (not very small) $\theta_{13}$ has already implied the presence of $CP$ violation! This is because the GR pattern we derive has a zero $\theta_{13}$ within LO approximation, and if we ignore the charged lepton contribution for $\theta_{13}$ due to its huge mass hierarchy(This assumption is reasonable since in the CKM quark mass mixing matrix, $\theta_{13}$ is significantly small due to its huge mass hierarchy.), the experimentally observed $\theta_{13}$ must come from $CP$ violation. On the other hand, our theory predicts $m_1=m_2$ within LO approximation, therefore the experimentally observed small mass splitting $\Delta m_{12}^2$ is also contributed by $CP$ violation. Interestingly, the current experimental results point to an approximate relation $|\Delta m_{12}^2/\Delta m_{23}^2|\sim \theta_{13}^2/\theta_{23}^2$. Since our topological scenario within LO approximation requires $\Delta m_{12}^2=0$ and $\theta_{13}=0$,
this might suggest that the nonzero $\Delta m_{12}^2$ and $\theta_{13}$ observed in current experiments might have a common origin -- the $CP$ violation.
We will leave a detailed study of $CP$ violation physics in our future publications.

\section{Conclusions and future directions}
In this paper, we start with a simple $1$D TSC model protected by $T^2=-1$ time reversal symmetry and show that the pair of time reversal protected topological Majorana zero modes on each end carries a $T^4=-1$ representation of time reversal symmetry and $1/4$ spin. Then we generalize the $T^4=-1$ fractionalized representation for a pair of topological Majorana zero modes into a $P^4=-1$ parity symmetry and a $\overline C^4=-1$ nontrivial charge conjugation symmetry as well. We also construct explicit condensed matter models and show that the proliferation of topological Majorana zero modes will lead to a relativistic dispersion and an $SU(2)$ spin.

These interesting observations from condensed matter systems motivate us to interpret a Majorana fermion as four topological Majorana zero modes(or a Lorentz spinon zero mode) and revisit its $\overline C,P,T$ symmetries. Surprisingly, we find that the $\overline C,P,T$ symmetries for a Majorana fermion made up of four Majorana zero modes satisfy a super algebra. The $\overline CPT$ super algebra for a Majorana fermion can be generalized into quantum filed theory as well. 
We further point out that the nontrivial charge conjugation symmetry $\overline C$ can be promoted to a $\mathbb{Z}_2$ gauge symmetry and its spontaneous breaking leads to the origin of (right-handed) neutrino mass. 
Indeed, the seesaw mechanism scenario requires such a fifth force.
In the classical limit, all the coupling terms arise from (gauge) interactions, hence it is not necessary to have any coupling term between SM particles and right-handed neutrinos since the right-handed sterile neutrino does not carry any SM gauge charge. However, in the seesaw mechanism, there is a coupling term in the form of $L \t\phi \nu_R$(the Dirac mass term with $L$ as the lepton doublets, $\phi$ as the Higgs field and $\nu_R$ as the right-handed neutrino field). Although it is an "allowed" term by gauge invariance, it is not a "natural" term since there is no interaction between $L \t\phi$ and $\nu_R$.
In the presence of the $\mathbb{Z}_2$ gauge force, such a term becomes natural since $\nu_L$ and $\nu_R$ can carry opposite half-$\mathbb{Z}_2$ charge. Here the concept of half-$\mathbb{Z}_2$ charge arises from the transformation Eq.(\ref{halfZ2}), where the $d_{L(R)}$ fermion operator takes eigen value $\mp i$ under charge conjugation symmetry, which is indeed a $\mathbb{Z}_4$ charge. The reason why a fermion can carry a half-$\mathbb{Z}_2$(or $\mathbb{Z}_4$) charge is again due to the group extension of the nontrivial charge conjugation symmetry $\overline C$ over the fermion parity symmetry that makes the total symmetry group to be $\mathbb{Z}_4$.
The half-$\mathbb{Z}_2$ charge assignment of a single fermion is also consistent with the fact that the mass term(a fermion bilinear) carries $\mathbb{Z}_2$ charge one.

These novel concepts even provide us a natural way to understand the origin of three generations of neutrinos, as there are three inequivalent ways to form a pair of complex fermions(with opposite spin polarizations) out of four topological Majorana zero modes, characterized by the $T^4=-1$, $(TP)^4=-1$ and $(T\overline C)^4=-1$ fractionalized symmetries that the complex fermions carry. This argument requires that a Majorana fermion is not a point-like particle and has an internal structure at cutoff scale. In the semiclassical limit, together with the $\mathbb{Z}_2$ gauge (minimal coupling) principle, we are able to uniquely determine the $\overline C,P,T$ invariant mass term and compute the neutrino mass mixing matrix with \emph{no} fitting parameters within LO approximation(without CP violation and charged lepton contribution). We obtain $\theta_{12}=31.7^\circ, \theta_{23}=45^\circ$ and $\theta_{13}=0^\circ$(the golden ratio pattern), which is consistent with the an $A_5$ flavor symmetry. We further predict an exact mass ratio for the three mass eigenstates with $m_1/m_3=m_2/m_3=3/\sqrt{5}$.

For future works, we would like to point out several interesting directions along this line of thinking:
(a) \emph{The quark CKM mass mixing matrix.} It is possible to use similar topological scenario to compute the quark CKM mass mixing matrix. However, a crucial difference in the quark CKM mass mixing matrix is the mass hierarchy problem, which leads to a significant suppressing of its mixing angles. It is important to understand the origin of quark mass hierarchy.
(b)\emph{The cutoff problem.} 
A challenging and deep way to deal with the cutoff problem is to develop a mathematical framework for quantum field theory in discrete space-time. If a fundamental theory has extremely strong quantum fluctuation at cutoff scale, the discrete structure would become crucial. The potential TQFT and UMT2C framework proposed in this paper are very promising future direction.
(c) \emph{Hidden super algebra for the SM.} From experimental point of view, to avoid fine-tuning, a super algebraic structure of the SM is demanded. Recent experimental results on the Higgs mass near $126GeV$\cite{higgsmass1,higgsmass2} point to a relation $M_{\rm{Higgs}}\simeq(M_u+M_d+M_c+M_s+M_t+M_b)/\sqrt{2}$(the Higgs boson mass is intrinsically close to the summation of six quark masses divided by $\sqrt{2}$). We also notice $M_t\simeq M_W+M_Z$(top quark mass is intrinsically close to the summation of $W$ and $Z$ boson masses) by pass. If the above two relations are not coincident, they must
be strong indications that SM might satisfy a hidden super algebra.
We note that these interesting mass relations are merely among the known fermions and bosons in the SM, therefore they can not be explained by any traditional super-symmetry.
Nevertheless, the topological Majorana zero modes might provide us a natural way to understand these relations, as they will generate degenerated states with different fermion parities.

\section{Acknowledgement}
Z-C Gu thank T.K. Ng's invitation for IAS Program on
 Topological Materials and Strongly Correlated Electronic Systems at HKUST, where the work was initiated, and Henry Tye, K.T. Law, Z-X Liu, T. Liu for helpful discussions on early results. Z-C Gu especially thank John Preskill, Alexei Kitaev, Neil Turok's encourages and delightful discussions for this work. Z-C Gu also thank X-G Wen, Y-S Wu, D. Gaiotto's suggestions for improving presentations, and his wife Y-F Ge's help on investigating experiment results. This work is supported by the Government of Canada through Industry Canada and by the Province of Ontario through the Ministry of Research and Innovation.

\begin{appendix}

\section{The mapping between Majorana fermion representation and Wyel fermion representation}
In this section, we will show how to map a real four component Majorana fermion representation to the standard Wyel fermion representation. Let us start from the real gamma matrix $\bar \gamma_0\equiv \gamma_5$ and $\bar \gamma_i\equiv \gamma_i$($i=1,2,3$), which is the real gamma matrix first proposed by E. Majorana. Let us introduce a unitary transformation:
\begin{eqnarray}
\t U=\left(
    \begin{array}{cc}
      1+\sigma_2 & -i(1-\sigma_2) \\
      i(1-\sigma_2) & 1+\sigma_2 \\
    \end{array}
  \right),
\end{eqnarray}
It is straightforward to verify that:
\begin{eqnarray}
\t \gamma_0=\t U(-i\bar\gamma^0)\t U^\dagger=\left(
    \begin{array}{cc}
      0 & 1 \\
      1 & 0 \\
    \end{array}
  \right),
\end{eqnarray}
and
\begin{eqnarray}
\t \gamma_i=\t U(-i\bar\gamma^i)\t U^\dagger=\left(
    \begin{array}{cc}
      0 & \sigma_i \\
      -\sigma_i & 0 \\
    \end{array}
  \right),
\end{eqnarray}
Clearly, $\t \gamma_0$ and $\t \gamma_i$ are the corresponding gamma matrix representation for Wyel fermion. Therefore, from a real four component Majorana fermion field, we can construct a complex two component Wyel fermion representation via $\Psi=U\psi$. More explicitly, in terms of real components $\psi_1,\psi_2,\psi_3,\psi_4$, we have:
\begin{eqnarray}
\Psi=\t U=\frac{1}{2}\left(
    \begin{array}{c}
     (\psi_1+\psi_4)-i(\psi_2+\psi_3)  \\ (\psi_2-\psi_3)+i(\psi_1-\psi_4) \\
      -(\psi_2-\psi_3)+i(\psi_1-\psi_4) \\ (\psi_1+\psi_4)+i(\psi_2+\psi_3)
    \end{array}
  \right)\equiv   \left(  \begin{array}{c}
     \chi  \\ -i\sigma_2 \chi^*
    \end{array}\right)\nonumber\\,
\end{eqnarray}
where the complex field $\chi$ is the complex two component Wyel fermion field.

\section{$\overline C,P,T$ symmetries in momentum space}
\label{App:momentum}
In this section, we will use a momentum space picture to describe the three generations of neutrinos. First, let us examine the $\overline C,P,T$ symmetry transformations of the Fourier modes $\gamma_\sigma(\v k)=\frac{1}{\sqrt{V}}\int d^3 x e^{-i\v k \cdot \v x}\gamma_\sigma(\v x)$ and
$\gamma_\sigma^\prime(\v k)=\frac{1}{\sqrt{V}}\int d^3 x e^{-i\v k \cdot \v x}\gamma_\sigma^\prime(\v x)$. It is straightforward to derive:
\begin{eqnarray}
\overline C \gamma_{\uparrow}(\v k) \overline C^{-1} &=&
-\gamma_{\downarrow}^\prime (\v k); \quad
\overline C \gamma_{\downarrow}(\v k) \overline C^{-1}= -\gamma_{\uparrow}^\prime(\v k); \nonumber\\
\overline C \gamma_{\uparrow}^\prime(\v k) \overline C^{-1} &=&
\gamma_{\downarrow}(\v k); \quad
\overline C \gamma_{\downarrow}^\prime(\v k) \overline C^{-1}= \gamma_{\uparrow}(\v k),
\end{eqnarray}
\begin{eqnarray}
P \gamma_{\uparrow}(\v k) P^{-1}&=&
-\gamma_{\uparrow}^\prime(-\v k); \quad
P \gamma_{\downarrow}(\v k) P^{-1}= \gamma_{\downarrow}^\prime(-\v k);\nonumber\\
P \gamma_{\uparrow}^\prime(\v k) P^{-1}&=&
\gamma_{\uparrow}(-\v k); \quad
P \gamma_{\downarrow}^\prime(\v k) P^{-1}= -\gamma_{\downarrow}(-\v k),
\end{eqnarray}
\begin{eqnarray}
T \gamma_{\uparrow}(\v k) T^{-1} &=&
-\gamma_{\downarrow}(-\v k); \quad
T \gamma_{\downarrow}(\v k) T^{-1}= \gamma_{\uparrow}(-\v k); \nonumber\\
T \gamma_{\uparrow}^\prime(\v k) T^{-1} &=&
-\gamma_{\downarrow}^\prime(-\v k); \quad
T \gamma_{\downarrow}^\prime(\v k) T^{-1}= \gamma_{\uparrow}^\prime(-\v k),
\end{eqnarray}
We note that the above transformation rules are also correct for Majorana spinon $\t\gamma_\sigma,\t\gamma_\sigma^\prime$ and $\hat\gamma_\sigma,\hat\gamma_\sigma^\prime$.

We can apply the similar argument to the emergence of three generations of Majorana fermions for their Fourier modes in momentum space as well.
\begin{eqnarray}
c_L(\v k)&=& \gamma_\up(\v k)+i \gamma_\down(\v k);\quad c_R(\v k)= \gamma_\up^\prime(\v k)-i \gamma_\down^\prime(\v k)\nonumber\\
f_L(\v k)&=& \t\gamma_\up(\v k)+i \t\gamma_\up^\prime(\v k);\quad f_R(\v k)=\t\gamma_\down(\v k)+i \t\gamma_\down^\prime(\v k)\nonumber\\
d_L(\v k)&=& \hat\gamma_\up(\v k)-i \hat\gamma_\down^\prime(\v k);\quad d_R(\v k)=\hat\gamma_\up^\prime(\v k)-i\hat \gamma_\down(\v k),\nonumber\\
\end{eqnarray}
Under $TP,T$ and $T \overline C$ symmetries, they transform as:
\begin{eqnarray}
T c_L(\v k) T^{-1}&=&-i c_L^\dagger(\v k); \nonumber\\ T c_R(\v k) T^{-1}&=&i c_R^\dagger(\v k) \nonumber\\
(T\overline C) f_L(\v k) (T \overline C)^{-1}&=&-i f_L^\dagger(\v k); \nonumber\\ (T \overline C) f_R(\v k) (T \overline C)^{-1}&=&i f_R^\dagger(\v k)\nonumber\\
(TP) d_L(\v k) (TP)^{-1}&=&-i d_L^\dagger(-\v k); \nonumber\\ (TP) d_R(\v k) (TP)^{-1}&=&i d_R^\dagger(-\v k),
\end{eqnarray}

The Hamiltonian of massless Majorana fermion has the following form in momentum space, e.g., for $\psi_d(x)$:
\begin{eqnarray}
\mathcal{H}_d=\frac{1}{4}\sum_{\v k}\psi^\dagger(\v k) \bar \gamma_0 \bar \gamma_i k_i \psi(\v k),
\end{eqnarray}
where $\psi(\v k)$ is the Fourier mode of $\psi(\v x)$, defined as $\psi(\v k)=\frac{1}{\sqrt{V}}\int d^3 x e^{-i\v k \cdot \v x}\psi(\v x)$. It is straightforward to verify that $\psi^\dagger(\v k)=\psi^t(-\v k)$.
If we assume the chiral basis has a spin polarization in the $y$-direction, we can fix the momentum to be $\v k=(0,k,0)$. Thus, we obtain:
\begin{widetext}
\begin{eqnarray}
\mathcal{H}_d=\frac{1}{4}\sum_{k}\left[k\hat\gamma_\up(-k)\hat\gamma_\up(k)-k\hat\gamma_\down(-k)\hat\gamma_\down(k)-k\hat\gamma_\up^\prime(-k)\hat\gamma_\up^\prime(k)
+k\hat\gamma_\down^\prime(-k)\hat\gamma_\down^\prime(k)\right]
\end{eqnarray}
In terms of the chiral fermion fields $d_L(\v k)$ and $d_R(\v k)$, we have:
\begin{eqnarray}
\mathcal{H}_d=\frac{1}{2}\sum_{k}\left[k d_L^\dagger(k) d_L(k)-k d_R^\dagger(k) d_R(k)\right]
\end{eqnarray}
For any given momentum $\v k$, we can define its positive energy mode as a left-handed neutrino and the negative energy mode as a right-handed antineutrino. However, we note that the zero energy mode $d_L(E=0)$ and $d_R(E=0)$ still transform as:
\begin{eqnarray}
(TP) d_L(E=0) (TP)^{-1}&=&-i d_L^\dagger(E=0); \quad (TP) d_R(E=0) (TP)^{-1}=i d_R^\dagger(E=0),
\end{eqnarray}
Thus, both of them carry the $(TP)^4=-1$ fractionalized symmetry. Furthermore, since the zero energy mode $d_{L(R)}$ transforms trivially under Lorentz symmetry, we can say that the vacuum effectively carries a $(TP)^4=-1$ fractionalized symmetry. Such an observation is pretty interesting, as traditional quantum field theory assume a unique vacuum that carries a trivial representation of $TP$ symmetry. The experimental consequence of such a fractionalized symmetry will be investigated in our future work.

For $c_{L(R)}$ and $f_{L(R)}$, their Hamiltonian in momentum space read:
\begin{eqnarray}
\mathcal{H}_{c(f)}=\frac{1}{4}\sum_{\v k}\psi^\dagger(\v k)  \gamma_0 \gamma_i k_i \psi(\v k),
\end{eqnarray}
If we assume the chiral basis has a spin polarization in the $z$-direction, we can fix the momentum to be $\v k=(0,0,k)$
In terms of the $c_{L(R)}$ and $f_{L(R)}$ fermion operators, we have:
\begin{eqnarray}
\mathcal{H}_d&=&\frac{1}{2}\sum_{k}\left[k c_L^\dagger(k) c_L^\dagger(-k)-k c_R^\dagger(k) c_R^\dagger(-k)+h.c.\right];\nonumber\\
\mathcal{H}_f&=&\frac{1}{2}\sum_{k}\left[k f_L^\dagger(k) f_L^\dagger(-k)-k f_R^\dagger(k) f_R^\dagger(-k)+h.c.\right],
\end{eqnarray}
In the Nambu basis, we obtain:
\begin{eqnarray}
\mathcal{H}_c&=&\frac{1}{2}\sum_{k}\left[\begin{array}{cc}
c_L^\dagger(k)+c_R^\dagger(k), & c_L(-k)-c_R(-k)
\end{array}\right]\left(
\begin{array}{cc}
  0 & k \\
    k & 0 \\
    \end{array}
    \right)\left[\begin{array}{c}
    c_L(k)+c_R(k) \\
    c_L^\dagger(-k)-c_R^\dagger(-k)
    \end{array}\right],\nonumber\\
\mathcal{H}_f&=&\frac{1}{2}\sum_{k}\left[\begin{array}{cc}
f_L^\dagger(k)-f_R^\dagger(k), & f_L(-k)+f_R(-k)
\end{array}\right]\left(
\begin{array}{cc}
  0 & k \\
    k & 0 \\
    \end{array}
    \right)\left[\begin{array}{c}
    f_L(k)-f_R(k) \\
    f_L^\dagger(-k)+f_R^\dagger(-k)
    \end{array}\right],
\end{eqnarray}
\end{widetext}
After diagonalizing the above two Hamiltonians, we can again define a positive mode corresponding to the left-handed neutrino and a negative energy mode corresponding to the right-handed antineutrino.

Similarly, the zero energy mode $c_{L(R)}(E=0)$ and $f_{L(R)}(E=0)$ transform as:
\begin{eqnarray}
T c_L(E=0) T^{-1}&=&-i c_L^\dagger(E=0); \nonumber\\ T c_R(E=0) T^{-1}&=&i c_R^\dagger(E=0),
\end{eqnarray}
and
\begin{eqnarray}
(T \overline C) f_L(E=0) (T \overline C)^{-1}&=&-i f_L^\dagger(E=0); \nonumber\\ (T \overline C) f_R(E=0) (T \overline C)^{-1}&=&i f_R^\dagger(E=0),
\end{eqnarray}
Therefore, we can say that the vacuum for for $c_{L(R)}$ and $f_{L(R)}$ fermions can effectively carry $T^4=-1$ and $(T \overline C)^4=-1$ fractionalized symmetries.

\end{appendix}

\bibliography{neutrino}
\end{document}